\documentclass[reqno]{amsart}
\usepackage[foot]{amsaddr}
\usepackage[margin=3cm]{geometry}
\usepackage{amsmath, amssymb,amsthm}
\usepackage[shortlabels]{enumitem}
\usepackage{mlmodern}
\DeclareSymbolFont{largesymbols}{OMX}{cmex}{m}{n} 
\usepackage{mathtools} 
\usepackage{stmaryrd} 
\usepackage[dvipsnames]{xcolor}
\usepackage{tikz}
\usepackage{upref} 
\usepackage[colorlinks]{hyperref}
\hypersetup{citecolor=blue,filecolor=blue,linkcolor=blue,urlcolor=navyblue}
\definecolor{navyblue}{rgb}{0.0, 0.0, 0.5}

\newtheorem{thm}{Theorem}[section]
\newtheorem{prop}[thm]{Proposition}
\newtheorem{lem}[thm]{Lemma}

\theoremstyle{definition}

\newtheorem*{claim*}{Claim}

\newcommand{\Z}{\mathbb{Z}}

\newcommand{\C}{\mathbb{C}}

\renewcommand{\Re}{\operatorname{\mathrm{Re}}}

\newcommand{\E}{\mathbf{E}}
\newcommand{\oneb}{\mathbf{1}}
\renewcommand{\P}{\mathbf{P}}

\newcommand{\Tr}{\operatorname{Tr}}

\newcommand{\hs}{\mathrm{hs}}
\newcommand{\id}{\operatorname{id}}

\newcommand{\sans}[1]{\mathsf{#1}}

\newcommand{\op}[1]{\operatorname{#1}}
\newcommand{\wt}{\widetilde}

\newcommand{\supp}{\op{supp}}

\newcommand{\Q}{\mathcal{Q}}
\newcommand{\Ex}{\sans{E}}
\newcommand{\Li}{\mathcal L}
\newcommand{\oU}{\mathcal U}

\newcommand\numberthis{\stepcounter{equation}\tag{\theequation}}
\numberwithin{equation}{section}

\begin{document}

\title[]{Optimal Frobenius light cones for power-law systems\\ in any dimension} 

\author{Laura Shou$^{1,2}$}
\author{An Le$^{1,2}$}
\author{Twesh Upadhyaya$^{1,2}$}
\author{T. C. Mooney$^{1,2}$}
\author{Alexey V. Gorshkov$^{1,2}$}

\address{\normalfont$^1$Joint Quantum Institute, NIST/University of Maryland, College Park, MD 20742, USA}
\address{\normalfont$^2$Joint Center for Quantum Information and Computer Science,
NIST/University of Maryland, College Park, MD, 20742, USA}

\begin{abstract}
We prove optimal Frobenius Lieb--Robinson bounds for power-law systems with power-law exponent $\alpha$ in any dimension $d$, in particular showing that the algebraic to linear light cone transition occurs at $\alpha=d+1$. The proof uses a spectral clipping method and concentration result to allow for running a multiscale operator-norm argument which produces the optimal Frobenius norm exponent. Additionally, we prove there is a Frobenius light cone transition from polylogarithmic to algebraic at the summability threshold $\alpha=d$, and a transition from decaying in volume to non-decaying in volume at $\alpha=d/2$.
\end{abstract}

\maketitle

\section{Introduction}

Light cones limit the spread of operators under evolution by spatially local Hamiltonians. The locality of the Hamiltonians' terms can be strict, as in the case of nearest-neighbour interactions, or effective, as in the case of power-law systems, where interactions exist between all sites but decay with distance $r$ as $1/r^\alpha$ for a specified power-law exponent $\alpha$ \cite{chen2023speed}.
Such light cones take the form of upper bounds on the growth of commutators between Heisenberg-evolved operators on distant sites. The bounds depend on the evolution time $t$ and on other parameters, such as $\alpha$.

Lieb--Robinson bounds \cite{lieb1972finite} constrain the operator norm of such commutators. They find application in bounding the time required for several tasks in quantum information, such as signaling, entangled state generation, and state transfer over perfectly or nearly perfectly initialized lattices \cite{Bravyi_2006_LRB,Nachtergaele_2006,tran2020hiearchy,chen2021optimal,Upadhyaya2026,Upadhyaya2026a}.

Beginning with early work by \cite{hastings2006spectral}, a bevy of recent results have completed the story of Lieb--Robinson bounds in power-law systems \cite{foss-feig2015nearly,guo2020signaling,chen2019finite,kuwahara2020strictly,tran2021lieb,tran2021optimal,chen2023speed}. Over regular lattices in all spatial dimensions $d$, tight bounds are now known for all $\alpha>d$. Here, tight means optimal scaling with distance up to polylogarithmic or subpolynomial factors. The light cone transitions from polylog to algebraic and from algebraic to linear are known to occur at $\alpha=2d$ and $2d+1$, respectively.

More recently, light cones bounding the \emph{Frobenius} norm of these commutators have been established. This norm averages over all singular values of the commutator, unlike the operator norm which only takes the largest singular value into account. Frobenius norms govern a different set of quantum information phenomena and tasks, like the onset of many-body quantum chaos and state transfer over uninitialized or nearly uninitialized lattices \cite{tran2020hiearchy,chen2021optimal,Upadhyaya2026,Upadhyaya2026a}. 

Remarkably, the Frobenius and operator-norm light cones can be qualitatively different in power-law systems, unlike for nearest-neighbor models. As compared to the usual operator-norm Lieb-Robinson bounds, the story of Frobenius-norm bounds in power-law systems has only just begun \cite{tran2020hiearchy,kuwahara2021absence,chen2021optimal}. Prior to this work, the only known tight bounds were in 1D for $\alpha>d$ \cite{tran2020hiearchy,chen2021optimal}.

We significantly extend this story, obtaining optimal Frobenius light cones in arbitrary dimension $d$, in particular locating the algebraic to linear Frobenius light cone transition at $\alpha=d+1$, resolving the conjectures from \cite{tran2020hiearchy,chen2021optimal,chen2023speed}. 
We also prove a logarithmic (fractional-power or polylogarithmic) to algebraic lightcone transition at $\alpha=d$, and a transition from times decaying in system size to non-decaying at $\alpha=d/2$, below the summability threshold.

At a technical level, the existing methods for deriving operator norms in power-law systems do not readily carry over to Frobenius-norm bounds. 
Our key new technical tools for the algebraic to linear transition bound are a spectral clipping argument, whereby the contribution of a small number of large eigenvalues can be controlled, combined with a multiscale argument typically used for operator norm Lieb--Robinson bounds in power-law systems \cite{tran2021lieb}. The effectiveness of such an argument relies on a concentration inequality from \cite{chen2021optimal}.
For the regime $\alpha\le d$, we prove Frobenius Lieb--Robinson bounds as well as fast protocol times.
Unlike the standard operator-norm Lieb--Robinson bounds and saturating protocols for $\alpha<d$, which decay in the system volume \cite{guo2020signaling,yin2025fast,shou2026fast}, we prove non-decaying Frobenius Lieb--Robinson bounds and polylogarithmic protocol times for $d/2<\alpha< d$, locating the fractional/polylogarithmic to algebraic light cone transition at $\alpha=d$. For $\alpha<d/2$, we obtain optimal (up to a logarithm) Frobenius bounds and protocols which decay in the system size. 

\subsection{Set-up}

We consider a Hamiltonian acting on a $d$-dimensional lattice $\Lambda$, which we may think of for simplicity as $\Z^d$, of qubits,
\begin{align}\label{eqn:H}
H(t)=\sum_{\{x,y\}\in\Lambda}h_{xy}(t)+\sum_{x\in\Lambda}h_x(t),\qquad\|h_{xy}(t)\|_\infty\le\frac{1}{|x-y|_\infty^\alpha},
\end{align}
for 2-site terms $h_{xy}(t)=\sum_{a,b=1}^3J_{xy}^{ab}(t)X_x^aX_y^b$ for $X_j^a$ Pauli operators, 1-site terms $h_x(t)=\sum_a h_x^a(t)X_x^a$ for $X_j^a$ identity or Pauli operators, and where $|x-y|_\infty=\max_{i}|x_i-y_i|$ for $x\ne y$. Note that we wrote this with $\ell^\infty$ norm since we use this norm throughout the paper, but it is equivalent to the usual Euclidean norm up to a $\sqrt{d}$ constant factor, which is absorbed into constants anyway. Note also that all the $2$-site terms are zero-trace, as the nonzero trace terms are incorporated into the single-site terms. 
For a finite set of sites $\Lambda$, the normalized Frobenius norm is
\begin{align}\label{eqn:frobenius-norm}
\|X\|_2^2:=\frac{\Tr_\Lambda[(X^\dagger X)]}{\Tr_\Lambda I}=2^{-|\Lambda|}\Tr_\Lambda(|X|^2).
\end{align}
For $H(t)$ as in \eqref{eqn:H}, and letting $A_x,B_y$ denote self-adjoint single-site operators with $\|A_x\|,\|B_y\|\le1$, define
\begin{align}\label{eqn:T2}
T_2(R):=\inf\{0<t<\infty:\exists\text{$H(t)$, and } A_x,B_y\text{ with $|x-y|_\infty\ge R$ and }\|[A_x(t),B_y]\|_2\ge1/2\},
\end{align}
for which, unless otherwise stated, we also allow arbitrarily large $\Lambda$ so we may consider $R\to\infty$.
Note the choice of $1/2$ is arbitrary; one could also choose a different constant between $0$ and $2$.
As is standard, throughout we consider a finite (but possibly large) region $\Lambda$ to make sense of the normalized Frobenius norm \eqref{eqn:frobenius-norm}. 
Unless otherwise stated, the constants in the bounds below are independent of the finite region $\Lambda$.

\subsection{Main results}
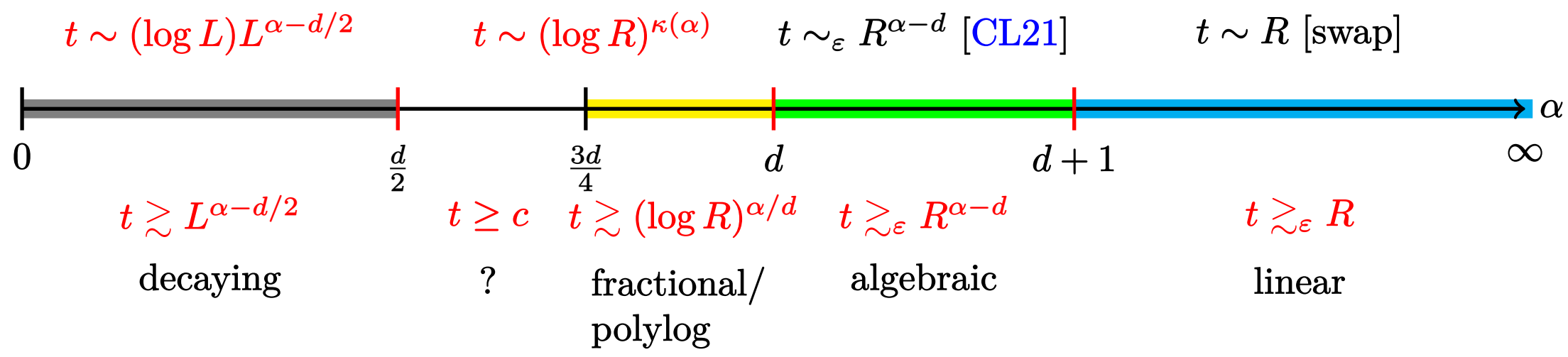
\begin{figure}[htb]
\begin{tikzpicture}[xscale=1.4]
\draw[line width=5,cyan] (7,0)--(10.05,0);
\draw[line width=5,green] (5,0)--(7,0);
\draw[line width=5,yellow] (3.75,0)--(5,0);
\draw[line width=5,gray] (0,0)--(2.5,0);

\begin{scope}[color=black, line width=1] 
\def\theight{-2mm};
\draw[->](0,0)--(10,0) node[right] {$\alpha$};
\draw (0,-\theight)--(0,\theight) node[below] {$0$};
\draw[red] (2.5,-\theight)--++(0,{2*\theight}) node[below,black] {$\frac d2$};
\draw (3.75,-\theight)--++(0,{2*\theight}) node[below] {$\frac{3d}{4}$};
\draw[red] (5,-\theight)--++(0,{2*\theight}) node[below,black] {$d$};
\draw[red] (7,-\theight)--++(0,{2*\theight}) node[below,black] {$d+1$};
\node[below] at (10,\theight) {$\infty$};
\end{scope}

\begin{scope}[yshift=-10mm,color=red] 
\node at (8.5,0) {$t\gtrsim_\varepsilon R$};
\node at (6,0) {$t\gtrsim_\varepsilon R^{\alpha-d}$};
\node at (4.4,0) {$t\gtrsim(\log R)^{\alpha/d}$};
\node at (3.1,0) {$t\ge c$};
\node at (1.25,0) {$t\gtrsim L^{\alpha-d/2}$};
\end{scope}

\begin{scope}[yshift=7mm] 
\node at (8.5,0) {$t\sim R$ [swap]};
\node at (6,0) {$t\sim_\varepsilon R^{\alpha-d}$ \cite{chen2021optimal}};
\node[red] at (3.8,0) {$t\sim(\log R)^{\kappa(\alpha)}$};
\node[red] at (1.25,0) {$t\sim (\log L) L^{\alpha-d/2}$};
\end{scope}

\begin{scope}[yshift=-1.6cm] 
\node at (8.5,0) {linear};
\node at (6,0) {algebraic};
\node[below] at (4.5,.25) {\parbox{2cm}{fractional/\\polylog}};
\node at (1.25,0) {decaying};
\node at (3.1,0) {?};
\end{scope}
\end{tikzpicture}
\caption{Frobenius norm protocols and bounds as a function of the power $\alpha$ (values at the boundaries not shown). 
The times above the axis represent the current fastest known protocols achieving large commutator in Frobenius norm for general dimension $d$, to the best of our knowledge. The bounds below the axis show Frobenius Lieb--Robinson lower bounds for general dimension $d$. New bounds and protocol times proved in this paper are shown in red. Our results rigorously locate Frobenius norm light cone transitions at $\alpha=d/2$, $d$, and $d+1$.
All scalings hide constant order factors, and notation with $\varepsilon$ indicates the scaling is up to a power $r^{\pm\varepsilon}$ for arbitrarily small $\varepsilon$. In the volume-dependent region below $d/2$, $L$ is the cube side-length.
}\label{fig:alpha}
\end{figure}

Our main results are depicted in Figure~\ref{fig:alpha}. We describe the results more precisely (including some of the bounds at the boundaries which are not depicted in the figure) in the following theorems.

We first consider the regime $\alpha>d$.
In $d=1$, \cite{chen2021optimal} proved a Frobenius algebraic to linear light cone transition at $\alpha=1+1$, and constructed protocols in any dimension which take time $t\sim r^{\op{min}\{\alpha-d,1\}+\varepsilon}$ for any $\varepsilon>0$. They conjectured that the light cone transition occurs in any dimension at $\alpha=d+1$, but their lower bounds did not extend to $d>1$.
Ref.~\cite{kuwahara2021absence} proved algebraic Frobenius lower bounds, but with non-optimal exponents, which do not locate the transition location.
Our first result proves the sharp Frobenius light cone exponent for $\alpha>d$ in any dimension, thus proving the conjectured $\alpha=d+1$ light cone transition. 

\begin{thm}[Frobenius lower bounds and light cone transition]\label{thm:frobenius}
Consider the Hamiltonian \eqref{eqn:H} on a $d$-dimensional lattice and with $\alpha>d$, and let $T_2(R)$ be the time defined in \eqref{eqn:T2}. Then for any $\varepsilon>0$ and sufficiently large $R\ge r_0$,
\begin{align}\label{eqn:t2-lower}
T_2(R)&\gtrsim_{\varepsilon,\alpha,d} R^{\mathrm{min}\{\alpha-d,1\}-\varepsilon}.
\end{align}
Combined with known protocols from \cite{chen2021optimal}, we thus obtain
\begin{align}\label{eqn:t2log}
\lim_{R\to\infty}\frac{\log T_2(R)}{\log R}=\mathrm{min}\{\alpha-d,1\},
\end{align}
which locates the algebraic to linear light cone transition at $\alpha=d+1$.
\end{thm}

Next, we consider the regime $\alpha\le d$, where the sum over interaction strengths $1/r^\alpha$ at a single site diverges in dimension $d$. Physically, the Hamiltonian is sometimes scaled by an overall Kac normalization factor in this regime to ensure the total energy does not blow up \cite{kac1963,defenu2023}. We omit this factor to remain consistency with existing Lieb--Robinson bound literature, but its reintroduction would simply rescale all times by that factor.
Despite the divergence already at $\alpha=d$, we show that we do not obtain a decaying or volume-dependent Frobenius Lieb--Robinson bound until the smaller power $\alpha\le d/2$. We have a constant lower bound for $d/2<\alpha\le 3d/4$, and a growing log-power lower bound for $\alpha>3d/4$. 
\begin{thm}[$\alpha\le d$]\label{thm:led}
Consider the Hamiltonian \eqref{eqn:H}.
For $\alpha\le d/2$, we consider $\Lambda$ to be a cube of side length $L$; otherwise the region for the lower bounds does not matter by square-summability of interaction terms. 
\begin{enumerate}[(i)]
\item There are the lower bounds 
\begin{align}\label{eqn:alphad-lrb}
T_2(R)\gtrsim_{\alpha,d}\begin{cases}
L^{\alpha-d/2},&0\le \alpha<d/2\\
1/\sqrt{\log L},&\alpha=d/2\\
1,&d/2<\alpha\le 3d/4\\
(\log R)^{\alpha/d},&3d/4<\alpha<d\\
\frac{\log R}{\log\log R},&\alpha=d
\end{cases}.
\end{align}

\item Additionally, there are protocols and single-site Paulis $A_x,B_y$ such that $\|[A_x(t),B_y]\|_2\ge1/2$ in time 
\begin{align}\label{eqn:alphad-protocol}
T\lesssim_{\alpha,d}\begin{cases}(\log L)L^{\alpha-d/2},&\alpha\ge0\\
(\log R)^{\frac{\alpha}{d}+\frac{\log 2}{\log(d/\alpha)}},& d/2<\alpha<d
\end{cases}.
\end{align}
\end{enumerate}
\end{thm}

Intuitively, one might expect the transition from non-decaying time to decaying time to indeed occur at $\alpha=d/2$ rather than at $d$ since the Frobenius norm involves the \emph{square-sum} of terms $\frac{1}{|x-y|_\infty^{2\alpha}}$, which diverges only for $\alpha\le d/2$. This transition location is also anticipated based on arguments for random chaotic systems in \cite{zhou2020operator, Zhou_2023}.

\subsection{Proof overview for Theorem~\ref{thm:frobenius}}
We briefly explain the proof ideas for Theorem~\ref{thm:frobenius}.
As discussed in \cite{chen2021optimal}, the multiscale argument for power-law Lieb--Robinson bounds in operator norm does not carry over cleanly to the normalized Frobenius norm. As a result, Frobenius light cones were typically studied using other methods specific to the Frobenius norm \cite{tran2020hiearchy,yin2020bound,chen2021optimal}. Ref.~\cite{chen2021optimal} showed how to adapt a multiscale argument to work with Frobenius norm in 1D, by proving a form of approximate near-submultiplicativity for the normalized Frobenius norm. This allowed for running a multiscale argument in Frobenius norm to obtain optimal light cones in 1D; however, the methods are specific to 1D and do not readily extend to higher dimensions $d$. 

\begin{figure}[htb]
\begin{tikzpicture}
\def\scolor{red}
\begin{scope}[scale=.35]
\def\gsize{4}
\def\osize{4.5}
\draw[\scolor,line width=1,fill=\scolor!50,opacity=.5] (-1.5,-1.5) rectangle (1.5,1.5);
\draw[dashed] (-\osize,-\osize) rectangle (\osize,\osize);
\node[\scolor,below] at (-.5,-1.5) {$Q$};
\node[below] at (0,-\osize) {$\op{supp}V_{j,Q}\subseteq Q^{+\ell_j}$};
\foreach \x in {-\gsize,...,\gsize}{
\foreach \y in {-\gsize,...,\gsize}{
\draw[fill=black] (\x,\y) circle (.7mm);}}
\end{scope}
\draw[->,yshift=.5cm](2,0)--++(.5,0);
\begin{scope}[xshift=5cm,yshift=.5cm]
\def\theight{1mm}
\def\sheight{4mm}
\def\rad{.6mm}
\def\sep{-1.4}
\foreach \row in {0,\sep}{
\draw[yshift={\row cm},->] (-2,0)--(2,0) node[right] {$\lambda$};
\draw[red,yshift={\row cm}] (-1,\theight)--(-1,-\theight) node[below] {$-\xi_j$};
\draw[red,yshift={\row cm}] (1,\theight)--(1,-\theight) node[below] {$\xi_j$};
\foreach \ev in {-7,...,7}{
\fill[yshift={\row cm}] (\ev/8,0) circle (\rad);
}
}
\foreach \ev in {-1.2,-1.5,1.3,1.5,1.8}{
\fill[red] (\ev,0) circle (\rad);
\ifdim \ev pt < 0 pt
    \fill[red,yshift={\sep cm}] (-1,0) circle (\rad);
    \draw[red,->] (\ev,0)--(-1,-1.2);
\else
    \fill[red,yshift={\sep cm}] (1,0) circle (\rad);
    \draw[red,->] (\ev,0)--(1,-1.2);
\fi
}
\node at (-2,.3) {$V_{j,Q}$};
\node[yshift={\sep cm}] at (-2,.3) {$G_{j,Q}$};
\end{scope}
\draw[->,yshift=-.9cm](7.5,0)--++(.5,0);
\begin{scope}[xshift=10cm,scale=.3]
\def\gsize{6}
\draw[gray] (-\gsize,-\gsize) grid (\gsize,\gsize);
\foreach \lj/\op in {1/90,3/60,\gsize/30}{
\fill[\scolor!\op,opacity=.3] (-\lj,-\lj) rectangle (\lj,\lj);
\draw[\scolor,line width=.7] (-\lj,-\lj) rectangle (\lj,\lj);
}
\fill[black] (0,0) circle (1mm);
\node at (-.4,-.4) {$x$};
\draw[->,yshift=.5cm,xshift=.5cm] (0,-1)--(0,-\gsize);
\begin{scope}[color=\scolor]
\node at (0,-1.25) {$\ell_1$};
\node at (0,-3) {$\ell_2$};
\node at (0,-\gsize) {$\ell_3$};
\end{scope}
\end{scope}
\end{tikzpicture}
\caption{
Depiction of the steps of the proof in the lower bound for $\alpha>d$. (Left) We first decompose $H(t)$ into a sum of Hamiltonians $V_{j,Q}$ based on spatial location $Q$ and interaction lengths $\ell_j$ [see \eqref{eqn:H-decomp}]. (Center) For each block $V_{j,Q}$, we clip large eigenvalues above a cutoff $\xi_j$ to form the clipped block $G_{j,Q}$. (Right) We use the clipped blocks $G_{j,Q}$ to form a new Hamiltonian $\wt H(t)$, and run the multiscale operator-norm argument with $\wt H(t)$ instead of $H(t)$.
}\label{fig:multiscale}
\end{figure}
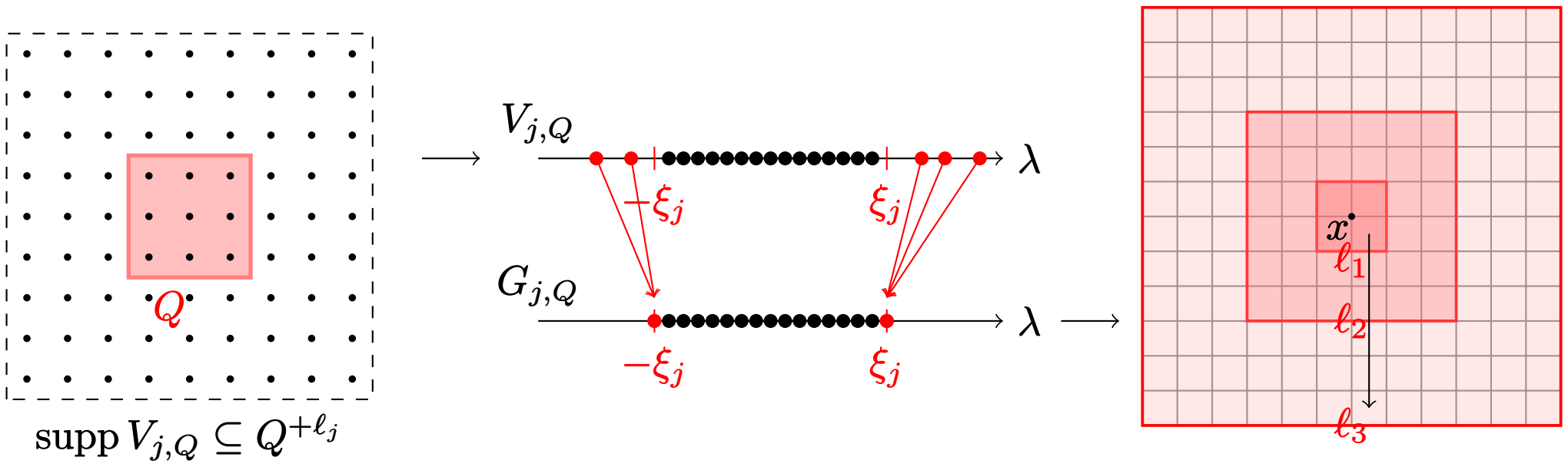

In this paper, we obtain optimal Frobenius Lieb--Robinson bounds for $\alpha>d$ in any dimension $d$ by instead working \emph{directly} with the multiscale operator norm Lieb--Robinson methods \cite{tran2021lieb}. 
We start by decomposing the Hamiltonian $H(t)$ by interaction length scales as is typical for multiscale arguments.
However, before applying the multiscale argument, we clip large eigenvalues in each block in the decomposition of $H(t)$ to form a new Hamiltonian $\wt H(t)$, whose blocks have dramatically reduced operator norm (Figure~\ref{fig:multiscale}). We then run the multiscale argument on $\wt H(t)$ to get tight operator norm bounds, using the reduced operator norm to get bounds which are much smaller than usual. A $p$-moment bound from \cite{chen2021optimal} will show the clipping procedure did not alter the Hamiltonian in Frobenius norm that much, and the sharp operator norm bounds for $\wt H(t)$ will then transfer to sharp Frobenius bounds for the original $H(t)$.

\subsection{Outline and notation}

The rest of the paper is organized as follows.

\begin{itemize}
\item In Section~\ref{sec:background}, we review previous results and background used for the proofs of the main results.

\item In Section~\ref{sec:lower}, we prove Theorem~\ref{thm:frobenius}, the Frobenius lower bounds for $\alpha>d$, deferring the proof of the multiscale Lemma~\ref{lem:multiscale} to the next section.

\item In Section~\ref{sec:lemmaproof}, we prove Lemma~\ref{lem:multiscale}, used in the proof of Theorem~\ref{thm:frobenius}.

\item In Section~\ref{sec:led-lower}, we prove Theorem~\ref{thm:led}(i) on Frobenius Lieb--Robinson bounds for $\alpha\le d$.

\item In Section~\ref{sec:protocol}, we prove Theorem~\ref{thm:led}(ii) on the protocol bounds for $\alpha\le d$.

\end{itemize}

Throughout this paper, generic constants $C,C',c,c',\ldots$ may change within or from line to line. We use $\lesssim$ and $\gtrsim$ to hide implicit constants, and $\sim$ to mean both $\lesssim$ and $\gtrsim$. 
To emphasize dependence on certain parameters, we may sometimes use notation such as $\lesssim_\varepsilon$ to indicate the implicit constant depends on $\varepsilon$, but when clear from context we may also simply use $\lesssim$. In particular, we view $d,\alpha,\varepsilon$ as constants and may not always specify dependence on them.
We also make use of standard asymptotic notation $o,O,\Theta,\Omega,\omega$.
The norm $\|\cdot\|_2$ is the normalized Frobenius norm defined in \eqref{eqn:frobenius-norm}, and the operator norm is denoted as $\|\cdot\|$ or sometimes $\|\cdot\|_\infty$.
For a set $S\subseteq\Lambda$, we let $S^{+\rho}$ denote $S$ expanded by $\rho$, i.e. $S^{+\rho}=\{x\in\Lambda :\exists y\in S\text{ such that }|x-y|_\infty\le\rho\}$.

\section{Background and preliminaries}\label{sec:background}

In this section, we review several useful results from the literature. 
For additional background, see e.g.~\cite{hastings2010,chen2023speed}.

We first state the moment estimate from \cite{chen2021optimal}, which we will use to bound the Frobenius-norm error after the clipping procedure.
\begin{lem}[moment estimate {\cite[Appendix A]{chen2021optimal}}]\label{lem:m-bound}
Consider a Hermitian operator $G = \sum_{\{x,y\}} G_{xy}$ with $\Tr_xG_{xy}=0$ for all $y$ and $\Tr_yG_{xy}=0$ for all $x$. Then letting $\|X\|_p^p:=\frac{\Tr_\Lambda[(X^\dagger X)^{p/2}]}{\Tr_\Lambda I}=2^{-|\Lambda|}\Tr_\Lambda(|X|^p)$, for every $p\ge2$,
\begin{align}
\|G\|_p\le 2(p-1)\left(\sum_{\{x,y\}} \| G_{xy}\|_{\infty}^2\right)^{1/2}.
\end{align}
\end{lem}

We will also need a standard finite-range Lieb--Robinson bound \cite{lieb1972finite,hastings2010,chen2023speed} for the multiscale argument.
\begin{lem}[standard Lieb--Robinson bound]\label{lem:lrb}

    Consider an operator $G(t) = \sum_{Z} G_Z(t)$ on the Hilbert space $(\C^q)^{\Lambda}$, where each $G_Z$ is a Hermitian operator which is supported on a set $Z\subseteq\Lambda$ satisfying $|Z| \leq k$ and diam$(Z) \leq l$. Define
    \begin{align*}
        \gamma = \sup_{t,x} \sum_{Z \ni x} \|G_Z(t)\|_\infty.
    \end{align*}
    We can think of $\gamma$ as the maximum interaction strength acting on a single site in our system. Let $U_{G}(t,0)$ be the time-evolution operator generated by $G$. Then for an operator $O$ supported on a set $S$, and an operator $B$ supported on a set that is at least distance $r$ away from $S$, we have:
    \begin{align}\label{eq:Standard_LRB}
        \|[U_G^{\dagger}(t,0) O U_{G}(t,0), B] \|_{\infty} \leq 2 |S| \|O\|_{\infty} \|B\|_\infty\exp{(2ek\gamma t - r/l)}.
    \end{align}
    This also implies:
    \begin{align}
        \| (\id - \Ex_{S^{r}}) U_G^{\dagger}(t,0) O U_{G}(t,0)  \| \leq C |S| \|O \|_{\infty} \exp \left( 2ek\gamma t - r/l\right).
    \end{align}
    with $S^r := \{x \in \Lambda: dist(x,S) \leq r \}$ and $\Ex_{S^r}$ denotes the orthogonal projection onto the set $S^r$ and some constant $C > 0$.

\end{lem}
Equation~\eqref{eq:Standard_LRB} follows immediately from \cite[Theorem 1]{hastings2010} if we choose their parameter $\mu = 1/l$ in our lemma. The second bound follows by writing $O(t)-\Ex_{S^{+r}}[O(t)]$ as an average over unitaries supported on $(S^{+r})^c$ and applying the first part.
\\

Finally, we state the standard Duhamel identity for bounding time evolution in terms of the generating Hamiltonians, which follows by differentiating $\mathcal U_{H_1}(s,0)\mathcal U_{H_0}(t,s)O$ as defined below in $s$. 
\begin{lem}[Duhamel]
Let $\mathcal U_H(t,s)O:= U_H^{-1}(t,s)OU_H(t,s)$ denote the time evolution of an operator $O$ under $H$, with the shorthand (and abuse of notation) $e^{\mathcal L_Ht}:= \mathcal U_H(t,0)$.
Then for Hamiltonians $H_0$ and $H_1$,
\begin{align}
e^{\Li_{H_1}t}O-e^{\Li_{H_0}t}O&=i\int_0^t e^{\Li_{H_1}s}[H_1(s)-H_0(s),\mathcal U_{H_0}(t,s)O]\,ds,
\end{align}
and thus
\begin{align}
\|e^{\Li_{H_1}t}O-e^{\Li_{H_0}t}O\|_*&\le 2t\|O\|\sup_{0\le s\le t}\|H_1(s)-H_0(s)\|_*,
\end{align}
for $\|\cdot\|_*$ either operator or Frobenius norm.
\end{lem}

\section{Frobenius lower bound}\label{sec:lower}

In this section, we prove the lower bound \eqref{eqn:t2-lower} in Theorem~\ref{thm:frobenius}.
As is usual, we bound the norm of the commutator $[A_x(t),B_y]$ by bounding how far an observable can spread in time $t$. For a region $S\subset\Lambda$, let $\Ex_{S}$ denote the orthogonal projection onto $S$, equivalently the normalized partial trace over $S^c$, i.e.~for an operator $O$ supported in $\Lambda$, $\Ex_S(O) := 2^{-|S^c|} \Tr_{S^c}(O) \otimes I_{S^c}$.
Let $e^{\Li_Ht}A_x$ denote the time-evolved operator under evolution by $H(t)$, i.e. $e^{\Li_Ht}A_x\equiv U^{-1}(t)A_xU(t)$ for $U(t)=\mathcal T\exp(-i\int_0^t H(s)\,ds)$ the time-evolution operator for $H(t)$, and let $B_r(x)$ be the $\ell^\infty$ ball of radius $r$ centered at $x$ (by equivalence of norms one could also use Euclidean norm with appropriate constant factor adjustments if desired).

We will prove the lower bound \eqref{eqn:t2-lower} of Theorem~\ref{thm:frobenius} by showing:
\begin{prop}\label{prop:main}
Let $0<\varepsilon<\op{min}\{\alpha-d,1\}$.
There are $C,r_0>0$ depending only on $d,\alpha,\varepsilon$, such that for every Hamiltonian \eqref{eqn:H}, $r\ge r_0$, and single-site operator $A_x$ with $\|A_x\|\le1$,
\begin{align}
\|(\id-\Ex_{B_r(x)})e^{\mathcal L_Ht}A_x\|_2\le Cr^{-\varepsilon/2},\quad\text{for all } t\le r^{\op{min}\{\alpha-d,1\}-\varepsilon}.
\end{align}
\end{prop}
This says that the time-evolved observable $A_x(t):=e^{\Li_Ht}A_x$ primarily acts only within\footnote{Note that $(\id-\Ex_{B_r(x)})$ is not equivalent to $\Ex_{B_r(x)^c}$, as the former considers all interactions extending outside the ball, while the latter projects onto operators supported only in $B_r(x)^c$.} the ball $B_r(x)$ up through times $r^{\op{min}\{\alpha-d,1\}-\varepsilon}$. For $R:=|x-y|_\infty$, $r\le R/2$, and $\|B_y\|\le1$, we see
\begin{align*}
\|[A_x(t),B_y]\|_2&=\|[(\id-\Ex_{B_r(x)})A_x(t),B_y]\|_2\le2\|(\id-\Ex_{B_r(x)})A_x(t)\|_2,
\end{align*}
using in the first equality that an operator supported in $B_r(x)$ commutes with $B_y$. Thus Proposition~\ref{prop:main} implies Theorem~\ref{thm:frobenius}.

The rest of this section and Section~\ref{sec:lemmaproof} prove Proposition~\ref{prop:main}.

\subsection{Multiscale decomposition and clipping}\label{subsec:clipping}
First, by moving into the interaction picture\footnote{If $H(t)=H_0(t)+V(t)$ with $H_0(t)$ the single-site terms, then the time evolution operator is $U(t)=U_0(t)U_I(t)$ with $U_0(t)=\otimes_{i\in\Lambda}u_i(t)$ a tensor product and $U_I(t)$ generated by $H':=\sum_{\{i,j\}}U_0^{-1}(t)h_{ij}(t)U_0(t)$, which satisfies the same 2-site power-law property as $V$. Then $e^{\Li_H t}A_x\equiv U(t)^{-1}A_xU(t)=U_I^{-1}(t)(U_0^{-1}(t)A_xU_0(t))U_I(t)$, and $\tilde A_x(t):=(U_0^{-1}(t)A_xU_0(t))$ is still a single-site operator with the same norm.}, we can consider Hamiltonians with only 2-site terms,
\begin{align}\label{eqn:H-2}
H(t)=\sum_{\{x,y\}\in\Lambda}h_{xy}(t),\quad h_{xy}(t)=\sum_{a,b=1}^3J_{xy}^{ab}(t)X_x^aX_y^b.
\end{align}

Let $\Omega:=B_{r}(x)\cap\Lambda$.
We decompose the Hamiltonian by interaction distances and lattice locations. For a constant $m$ to be chosen later (depending only on $d,\alpha,\varepsilon$, and independent of $r$), define length-scales $\ell_j:=\lceil r^{j/m}\rceil$, $j=1,\ldots,m-1$. Also let $\Q_j$, $j=2,\ldots,m-1$, partition $\Omega$ into disjoint cubes with side lengths $\ell_j$. Let $V_1:=\sum_{\{x,y\}\subset\Omega:|x-y|_\infty<\ell_1}h_{xy}$, and write
\begin{align}\label{eqn:H-decomp}
H_{\Omega,\ell_{m-1}}=V_1+\sum_{j=2}^{m-1}\sum_{Q\in\Q_j}V_{j,Q},\quad\text{for}\quad V_{j,Q}:=\sum_{\substack{(x,y)\in Q\times \Omega, x\prec y,\\\ell_{j-1}\leq|x-y|_\infty<\ell_j}}h_{xy},
\end{align}
where $x\prec y$ denotes lexicographical ordering. Each $V_{j, Q}$ is supported in $Q^{+\ell_j}$.
Note that $H_{\Omega,\ell_{m-1}}$ is then the restriction of $H$ to $\Omega$ and with only interactions of length $<\ell_{m-1}$.

Let $f_\xi(\lambda):=\op{sgn}(\lambda)\min(|\lambda|,\xi)$, which clips $\lambda$ with large $|\lambda|>\xi$ to the nearest cutoff $\pm\xi$.
Writing $V_{j,Q}$ from \eqref{eqn:H-decomp} in its eigenbasis as $V_{j,Q}=\sum_\lambda \lambda|\lambda\rangle\langle\lambda|$, define for $\xi_j$ to be chosen later,
\begin{align}
G_{j,Q}&:=f_{\xi_j}(V_{j,Q})=\sum_\lambda\op{sgn}(\lambda)\min(|\lambda|,\xi_j)|\lambda\rangle\langle\lambda|,
\end{align}
which is a spectrally truncated or ``clipped'' version of $V_{j,Q}$. By construction, $\|G_{j,Q}\|_\infty\le \xi_j$, and for $p\ge2$,
\begin{align*}
\|V_{j,Q}-G_{j,Q}\|_2^2&=\frac{1}{2^{|\Lambda|}}\sum_\lambda(|\lambda|-\xi_j)_+^2
\le\frac{1}{2^{|\Lambda|}}\sum_\lambda |\lambda|^p\xi_j^{2-p}
= \xi_j^{2-p} \|V_{j,Q}\|_p^p.\numberthis\label{eqn:f-diff}
\end{align*}
Note in the intermediate inequality, taking large $p$ allows for a better approximation of the cutoff in $(|\lambda|-\xi_j)_+^2\le|\lambda|^2\oneb_{|\lambda|\ge\xi_j}$; the term $|\lambda|^p\xi_j^{-p}$ is amplified by large $p$ for $|\lambda|>\xi_j$, and strongly suppressed for $|\lambda|<\xi_j$.
The moment bound Lemma~\ref{lem:m-bound} will be used to show that the above Frobenius error \eqref{eqn:f-diff} is small.
The lemma and the power-law interactions give
\begin{align*}
\|V_{j,Q}\|_p\le 2(p-1)\Bigg(\sum_{\substack{(x,y)\in Q\times \Omega, x\prec y,\\\ell_{j-1}\leq|x-y|_\infty<\ell_j}}\|h_{xy}\|_\infty^2\Bigg)^{1/2}
&\le  2(p-1)\left(C\ell_j^d\sum_{y:|x-y|\ge\ell_{j-1}}|x-y|_\infty^{-2\alpha}\right)^{1/2}\\
&\le C (p-1)\ell_j^{d/2}\ell_{j-1}^{d/2-\alpha}.\numberthis\label{eqn:Vq-moment}
\end{align*}
Taking the cutoff $\xi_j:=2C(p-1)\ell_j^{d/2}\ell_{j-1}^{d/2-\alpha}$ which is larger than this bound, then \eqref{eqn:f-diff} and \eqref{eqn:Vq-moment} imply
\begin{align}\label{eqn:VG-frobenius}
\|V_{j,Q}-G_{j,Q}\|_2^2&\le \xi_j^2 2^{-p}.
\end{align}
Note for any $p=O(r)$, the cutoff $\xi_j$ is only $r^{O(1)}$ for any $j=1,\ldots,m-1$.

We now form a new Hamiltonian $\wt H(t)$ from the clipped blocks $G_{j,Q}$, as
\begin{align}\label{eqn:tildeH}
\wt H:=V_1+\sum_{j=2}^{m-1}\sum_{Q\in\Q_j}G_{j,Q},\quad\text{where}\quad G_{j,Q}=f_{\xi_j}(V_{j,Q}).
\end{align}
By \eqref{eqn:VG-frobenius}, taking $p:=\kappa\log r$ for sufficiently large $\kappa=\kappa(a',d)$ to counteract the sums over $j$ and $\mathcal Q_j$ and powers of $r$ in the $\xi_j^2$, we see $\wt H$ approximates $H_{\Omega,\ell_{m-1}}$ in Frobenius norm as
\begin{align}\label{eqn:frobenius-moment}
\|H_{\Omega,\ell_{m-1}}-\wt H\|_2&\lesssim r^{-a'}.
\end{align}
We also let $\wt H_k:=V_1+\sum_{j=2}^{k}\sum_{Q\in\Q_j}G_{j,Q}$. 
Note that each $G_{j,Q}$ has the same support as $V_{j,Q}$, since if we write $V_{j,Q}=V_S\otimes I_{S^c}$, then $f_\xi(V_{j,Q})=f_\xi(V_S)\otimes I_{S^c}$.

\subsection{Adding scales}
We now run the multiscale argument similar to \cite[Lemma S1]{tran2021lieb}, but applied to $\wt H(t)$ and streamlined. Since all the blocks in the decomposition \eqref{eqn:tildeH} of $\wt H(t)$ have dramatically reduced operator norm, the resulting bounds (in operator norm) will be much smaller than usual, and in particular will be good enough to obtain a sharp Frobenius norm.

We first consider $\wt H_1=V_1$, which contains interactions with range $<\ell_1$, and apply the standard Lieb--Robinson bound Lemma~\ref{lem:lrb} to $V_1$. 
Each term $h_{xy}$ in $V_1$ is supported on $2$ sites of distance $\le\ell_1$ apart, and since $\alpha>d$,
\begin{align}
\gamma=\sup_{t,x}\sum_{y:y\ne x}\|h_{xy}(t)\|&\le \sum_y\frac1{|x-y|_\infty^\alpha}\sim \int_1^\infty r^{-\alpha+d-1}\,dr=O(1).
\end{align}
Let $\rho:=4e\gamma\ell_1t+b_1$ for $b_1$ to be chosen later.
Then for an observable $O$ supported in $S\subseteq\Omega$ and satisfying $\|O\|\le1$, we get
\begin{align*}
\|e^{\Li_{V_1}t}O-\Ex_{S^{+\rho}}e^{\Li_{V_1}t}O\|&\le 2|S|\exp(2ek\gamma t-\rho/\ell_1)\\
&\le 2|S|\exp(-b_1/\ell_1)\le r^{-a},\numberthis\label{eqn:lrb1}
\end{align*}
by choosing $b_1=C_a\ell_1\log r$ with sufficiently large $C_a$ depending on the chosen power $a$ to counteract $|S|\le Cr^d$. 

Next, starting from $\wt H_k$, we bound how far observables can spread when we add the next scale from $\wt H_{k+1}$; compare \cite[Lemma S1]{tran2021lieb}. We use and prove a simplified/streamlined version of the multiscale argument compared to \cite{tran2021lieb}, though the idea (iteratively coarse-grain and apply the finite-range Lieb--Robinson bound at different scales) remains the same.
\begin{lem}[adding a scale]\label{lem:multiscale}
Fix a time cut-off $T\le r$ and consider a region $\Omega\subseteq\Lambda$ with $|\Omega|\le Cr^d$. Suppose that for some family of Hamiltonians $F_0(t)$ on $\Omega$, there exists some $b\leq \ell,$ $0\leq v\leq r^2,$ and $\eta>0$ such that for $A$ equal to either $F_0(t)$ or its inverse, $-F_0(T-t),$ the following holds for any interval of length $t$ in $[0,T]$, with $e^{\Li_At}$ denoting the propagator starting from the start of the interval:
\begin{align}\label{eqn:lrb-input}
\|e^{\Li_{A}t}O_S-\Ex_{S^{+(vt+b)}}e^{\Li_{A}t}O_S\|\le \eta\|O_S\|,
\end{align} for any observable $O_S$ supported in $S\subseteq \Omega.$ 

Let $H(t)=F_0(t)+G(t)$ for $G(t)=\sum_{Q\in\Q}G_Q(t)$ acting on $\Omega$, where $\Q$ is a partition of $\Omega$ consisting of $\ell$-cubes (i.e., cubes of side-length $\ell$), and each $G_Q(t)$ is supported in an $O(\ell)$-cube $S_Q\subseteq Q^{+C\ell}$ through time $T$ and satisfies $\|G_Q(t)\|\le g\le r^2$.
Then for any constant $a>0,$ and all intervals of length $t$ in $[0,T]$, for $B$ equal to $H(t)$ or its time-reversal $-H(T-t)$:
\begin{align}
\|e^{\Li_{B}t}O_S-\Ex_{S^{+(v't+b')}}e^{\Li_{B}t}O_S\|\le \eta'\|O_S\|,
\end{align}
with
\begin{align}\label{eqn:recursion}
b'\lesssim_a\ell\log r, \quad v'\lesssim_a v\log r+\ell g,\quad \eta'\lesssim r^{d+3}(\eta+r^{-a}).
\end{align}
\end{lem}
We give the proof of this lemma in Section~\ref{sec:lemmaproof}.

The initial scale is the estimate \eqref{eqn:lrb1} for $\wt H_1=V_1$, with $v_1=4e\gamma\ell_1\le C\ell_1$, $b_1=C_a\ell_1\log r$, and $\eta=r^{-a}$.
We now apply Lemma~\ref{lem:multiscale} to $\wt H_2,\wt H_3,\ldots,\wt H_{m-1}$ one at a time, using the previous output as the new initial scale \eqref{eqn:lrb-input}.
Note that $m$ to be chosen is a constant [see \eqref{eqn:m}], so accumulation of implicit constants from the lemma is not a concern. 
We have $\ell_k/\ell_{k-1}\sim r^{1/m}\gg \log r$, so for sufficiently large $r$ the condition $b_{j-1}\le\ell_j$ for Lemma~\ref{lem:multiscale} is always met. The support of $G_{j,Q}$ and $V_{j,Q}$ is contained in $S_Q=Q^{+\ell_j}$.
Recall for the cutoff $\xi_j$ and $p$ defined in Section~\ref{subsec:clipping}, we have the operator norm estimate
\begin{align}
\|G_{j,Q}\|&\le \xi_j\lesssim (\log r)\ell_j^{d/2}\ell_{j-1}^{d/2-\alpha}\lesssim (\log r)r^{\frac{j}{m}(d-\alpha)-\frac{d}{2m}+\frac{\alpha}{m}}.
\end{align}
We will later take $(\alpha-d/2)/m\le\varepsilon/2\le2$ [see \eqref{eqn:m}], so we see the $g\le r^2$ condition in the lemma is met.

Using the velocity recursion in \eqref{eqn:recursion}
\begin{align}
v_{k+1}\lesssim_a (\log r)v_k+\ell_{k+1}g_{k+1},
\end{align}
and letting $\beta:=\alpha-d$, we thus obtain
\begin{align*}
v_k&\lesssim_a [(\log r)]^{k-1}v_1+\sum_{j=1}^{k-1}r^{(1-\beta)_++(d/2+\beta)/m}(\log r)^{j}\\
&\lesssim_{a,m} (\log r)^mr^{(1-\beta)_++(d/2+\beta)/m}.\numberthis
\end{align*}
We take $m$ a constant depending only on $d,\beta,\varepsilon$ so that 
\begin{align}\label{eqn:m}
(d/2+\beta)/m\le \varepsilon/2.
\end{align}
Also, $b_{m-1}\lesssim r^{1-1/m}\log r=o(r)$.
Then in time $T=r^{\min\{\beta,1\}-\varepsilon}$, an observable evolved under $\wt H(t)=\wt H_{m-1}(t)$ can spread distance only order
\begin{align}
\nonumber v_{m-1}T+b_{m-1}&\lesssim (\log r)^m r^{(1-\beta)_++\min\{\beta,1\}+(d/2+\beta)/m-\varepsilon}+o(r)\\
&\lesssim (\log r)^m r^{1-\varepsilon/2}+o(r)=o(r).
\end{align}
More precisely, for $r\ge r_0$ and a single-site observable $A_x$ with $\|A_x\|\le1$, we obtain, from repeated application of Lemma~\ref{lem:multiscale}, that
\begin{align}\label{eqn:htilde-final}
\|e^{\Li_{\wt H}t}A_x-\Ex_{B_{r/2}(x)}e^{\Li_{\wt H}t}A_x\|&\le Cr^{-a'},
\end{align}
by choosing $a$ (to be fully determined later) depending on $d,m$ sufficiently large.
Additionally, the above holds for with the propagator $\mathcal U(t,s)$, for example just by considering a Hamiltonian starting at time $s$.

\subsection{Error terms}
It remains to relate $\wt H$ back to $H$. 
As usual, Duhamel bounds the time evolutions as
\begin{align}\label{eqn:duhamel-wt}
\|e^{\Li_Ht}A_x-e^{\Li_{\wt H}t}A_x\|_2&\le \int_0^t\|[H(s)-\wt H(s),\mathcal{U}_{\tilde{H}}(t,s)A_x]\|_2\,ds.
\end{align}
We will decompose $H=H_{\Omega,\ell_{m-1}}+H_{\Omega,\ge\ell_{m-1}}+H_{\partial\Omega,{\ell_{m-1}}}+H_{\Omega^c}$, the latter three terms defined as used below.
Letting $H_{\Omega^c}$ consist of all interactions supported entirely outside $\Omega$, we see $[H_{\Omega^c},\mathcal{U}_{\tilde{H}}(t,s)A_x]=0$ since $\wt H$ consists only of interactions inside $\Omega$. Thus for bounding \eqref{eqn:duhamel-wt}, we can ignore interactions entirely outside $\Omega$.
We then have 3 sources of errors: (1) the clipping error, (2) interactions of long length $\ge \ell_{m-1}$ touching $\Omega$, and (3) interactions of length $<\ell_{m-1}$ across the boundary of $\Omega$.
These types of errors in $H(s)-\wt H(s)$ are bounded as follows.
\begin{enumerate}
\item For the clipping error, we use $\|[D,O]\|_2\le 2\|D\|_2\|O\|$ with $D=\wt H(s)-H_{\Omega,\ell_{m-1}}(s)$ and the Frobenius bound \eqref{eqn:frobenius-moment}, giving
\begin{align}\label{eqn:error-1}
\|[\wt H(s)-H_{\Omega,\ell_{m-1}}(s),\mathcal{U}_{\tilde{H}}(t,s)A_x]\|_2&\lesssim 2r^{-a'}.
\end{align}

\item For the long range $\ge\ell_{m-1}$ interactions, we use the same commutator norm bound as in (1), with $D=H_{\Omega,\ge\ell_{m-1}}:=\sum_{\substack{\{x,y\}\subset\Lambda\\\{x,y\}\cap\Omega\ne\emptyset\\|x-y|_\infty\ge\ell_{m-1}}}h_{xy}$. Since $\Tr(h_{xy}^\dagger h_{uv})=0$ for $\{x,y\}\ne\{u,v\}$ for 2-site terms \eqref{eqn:H-2} by tracing over a site appearing only once, we obtain the Frobenius/Hilbert--Schmidt estimate
\begin{align}\label{eqn:error-2}
\|H_{\Omega,\ge\ell_{m-1}}\|_2^2&\le \sum_{x\in\Omega}\sum_{y:|x-y|_\infty\ge\ell_{m-1}}\frac{1}{|x-y|_\infty^{2\alpha}}\le C_{\alpha,d}r^d \ell_{m-1}^{d-2\alpha}\lesssim r^{-2(\alpha-d)+2(\alpha-d/2)/m}.
\end{align}

\item For $H_{\partial\Omega,\ell_{m-1}}=\sum_{\substack{x\in\Omega,y\not\in\Omega\\|x-y|_\infty<\ell_{m-1}}}h_{xy}(t)$ containing interactions of length $<\ell_{m-1}$ across the boundary of $\Omega$, note that $\ell_{m-1}=\lceil r^{(m-1)/m}\rceil=o(r)$. So for sufficiently large $r$, these interactions do not touch $\Ex_{B_{r/2}(x)}\mathcal U(t,s)A_x$ from \eqref{eqn:htilde-final}, and so $H_{\partial\Omega,\ell_{m-1}}$ commutes with $\Ex_{B_{r/2}(x)}\mathcal U(t,s)A_x$. Since $\alpha>d$, we can use summability over $y$ to see $\|H_{\partial\Omega,\ell_{m-1}}\|\le Cr^d$. Then with \eqref{eqn:htilde-final}, we obtain for $r\ge r_0$,
\begin{align}\label{eqn:error-3}
\|[H_{\partial\Omega,\ell_{m-1}},\mathcal U_{\tilde H}(t,s)A_x]\|_2&\lesssim r^{d-a'}.
\end{align}
\end{enumerate}
In total, we obtain
\begin{align}\label{eqn:error-final}
\|e^{\Li_H t}A_x-e^{\Li_{\wt H}t}A_x\|_2&\lesssim t(r^{-a'}+r^{-\beta+(d/2+\beta)/m}+r^{d-a'}).
\end{align}
For $t\le r^{\op{min}\{\beta,1\}-\varepsilon}$ and recalling $m$ was chosen in \eqref{eqn:m} so that $(d/2+\beta)/m\le \varepsilon/2$, the middle term's exponent is $\le-\varepsilon/2$. 
By choosing $a'$ large enough, combining \eqref{eqn:error-final} (applied twice, and using that partial trace can only reduce the norm) and \eqref{eqn:htilde-final} gives
\begin{align}
\|e^{\Li_Ht}A_x-\Ex_{B_{r/2}(x)}e^{\Li_{H}t}A_x\|_2&\le Cr^{-\varepsilon/2},
\end{align}
which with another triangle inequality estimate and noting $\Ex_{B_r}\Ex_{B_{r/2}}=\Ex_{B_{r/2}}$ to expand to $B_r(x)$ [cf.~\eqref{eqn:multiscale-initial2}], proves Proposition~\ref{prop:main}. \qed

\section{Proof of Lemma~\ref{lem:multiscale} multiscale argument}\label{sec:lemmaproof}
In this section, we prove Lemma~\ref{lem:multiscale} on adding scales in the multiscale argument.

\begin{proof}[Proof of Lemma~\ref{lem:multiscale}]
Let $U_0$ be the propagator for $F_0$.
We work in the interaction picture of $F_0$, so a single $G_Q$ transforms as
\begin{align*}
W_Q(t):=U_0(t)^{-1}G_Q(t)U_0(t)\equiv e^{\Li_{F_0}t}G_Q(t),
\end{align*}
where $U_0(t)=\mathcal T\exp(-i\int_0^t F_0(s)\,ds)$.
The propagator $U$ for $H$ factors as $U(t)=U_0(t)U_W(t)$, for $U_W$ capturing time-evolution due to the perturbation $W(t)$, and $e^{\Li_Ht}=e^{\Li_Wt}e^{\Li_0t}$. We first use \eqref{eqn:lrb-input} to consider how $G_Q$ spreads under $F_0$, then use the finite-range Lieb--Robinson bound Lemma~\ref{lem:lrb} to consider spreading due to the interaction $G=\sum_Q G_Q$.

In order to control the time evolution using \eqref{eqn:lrb-input}, we start by considering small times $t\le \ell/v$, so that $vt+b\le 2\ell$.
The hypothesis \eqref{eqn:lrb-input} then implies
\begin{align*}
\|W_Q(t)-\Ex_{S_Q^{+2\ell}}W_Q(t)\|&\le \|W_Q(t)-\Ex_{S_Q^{+(vt+b)}}W_Q(t)\|+\|\Ex_{S_Q^{+2\ell}}(\Ex_{S_Q^{+(vt+b)}}W_Q(t)-W_Q(t))\|\\
&\le 2\eta g,\numberthis\label{eqn:multiscale-initial2}
\end{align*}
using that $\|W_Q\|=\|G_Q\|\le g$ and that partial trace can only decrease the norm.
Thus $W_Q(t)$ is close to the operator $W'_Q(t):=\Ex_{S_Q^{+2\ell}}W_Q(t)$ which has support in $S_Q^{+2\ell}$.
Consider the full perturbations $W(t):=\sum_Q W_Q(t)$ and $W'(t):=\sum_Q W'_Q(t)$.
Since the number of cubes $Q$ is $\le Cr^d$, we see for any $O$,
\begin{align*}
\|[W(t)-W'(t),O]\| &\le 2\|W(t)-W'(t)\|\|O\|\le 4Cr^dg\eta\|O\|.
\end{align*}
Applying Duhamel then gives for any $O$,
\begin{align}\label{eqn:W-Wprime}
\|e^{\Li_Wt}O - e^{\Li_{W'}t}O \|&\le 4Cr^d g\eta t\|O\|.
\end{align}

We can now consider evolution due to $W$. From \eqref{eqn:W-Wprime}, we can pay the above error to instead work with $W'(t)$, whose individual summands have supports in regions $S_Q^{+2\ell}$. The standard Lieb--Robinson bound Lemma~\ref{lem:lrb} does not depend on the local Hilbert space dimension, so we can apply it to $W'$ with $\ell$-cubes $Q$ as coarse lattice sites.
We have $\|W_Q'\|\le \|W_Q\|\le g$, and $Z=\op{supp}W_Q'\subseteq Q^{+(C+2)\ell}$. In terms of lattice units $\ell$, $\op{diam}Z=O(1)$. Also, $Z$ intersects at most $C=C(d)$ other ``points'' $Q'$, so $k=O(1)$. Similarly, for any $\ell$-cube $B$, there are only $O(1)$ terms $W_{Q}'$ which can interact with $B$ (since the range of $W_Q'$ is $\le(C+2)\ell$ and $Q$ has size $\ell$), so $\gamma=O(g)$.
Thus for $\rho=2ek\gamma|t|\ell+C_a\ell\log r\sim C\ell g|t|+C_a\ell\log r$, with sufficiently large $C_a$, the standard Lieb--Robinson bound Lemma~\ref{lem:lrb} gives
\begin{align}\label{eqn:scale-lrb2}
\|e^{\Li_{W'}t}O_S-\Ex_{S^{+\rho}}e^{\Li_{W'}t}O_S\|&\le Cr^{-a}\|O_S\|.
\end{align}

Finally, we undo the interaction picture and combine the estimates \eqref{eqn:W-Wprime} and \eqref{eqn:scale-lrb2}. Let $H'=F_0+G'$ be the Hamiltonian whose interaction-picture perturbation is $W'$, i.e. $U_0^{-1}(t)G'U_0(t)=W'(t)$; then
\begin{align*}
\|e^{\Li_Ht}O_S-\Ex_{S^{+(\rho+vt+b)}}[e^{\Li_Ht}O_S]\|&\le \begin{multlined}[t]\|e^{\Li_Ht}O_S-e^{\Li_{F_0+G'}t}O_S\|\\
+\|e^{\Li_{F_0+G'}t}O_S-\Ex_{S^{+(\rho+vt+b)}}[e^{\Li_{F_0+G'}t}O_S]\|\\
\qquad\qquad+\|\Ex_{S^{+(\rho+vt+b)}}[e^{\Li_{F_0+G'}t}O_S]-\Ex_{S^{+(\rho+vt+b)}}[e^{\Li_{H}t}O_S]\|.
\end{multlined}\numberthis
\end{align*}
Since $e^{\Li_Ht}=e^{\Li_W t}e^{\Li_0t}$ and $e^{\Li_{H'}t}=e^{\Li_{W'}t}e^{\Li_0t}$, the first and third terms above are bounded using \eqref{eqn:W-Wprime}, and the second using \eqref{eqn:scale-lrb2} and \eqref{eqn:lrb-input} as
\begin{align*}
\|e^{\Li_{F_0+G'}t}O_S-\Ex_{S^{+(\rho+vt+b)}}&[e^{\Li_{F_0+G'}t}O_S]\|\\
&\le\begin{multlined}[t] \|e^{\Li_{W'}t}e^{\Li_{0}t}O_S-e^{\Li_{W'}t}\Ex_{S^{+(vt+b)}}[e^{\Li_0t}O_S]\|\\
+\|e^{\Li_{W'}t}\Ex_{S^{+(vt+b)}}[e^{\Li_0t}O_S]-\Ex_{S^{+(\rho+vt+b)}}[e^{\Li_{W'}t}\Ex_{S^{+(vt+b)}}[e^{\Li_0t}O_S]]\|\\
+\|\Ex_{S^{+(\rho+vt+b)}}[e^{\Li_{W'}t}\Ex_{S^{+(vt+b)}}[e^{\Li_0t}O_S]]-\Ex_{S^{+(\rho+vt+b)}}[e^{\Li_{W'}t}e^{\Li_0t}O_S]\|
\end{multlined}\\
&\le (Cr^{-a}+2\eta)\|O_S\|. \numberthis
\end{align*}
Thus for short times $t\le \ell/v$ and $\|O_S\|\le1$, we obtain
\begin{align}\label{eqn:short-error}
\|e^{\Li_{H}t}O_S-\Ex_{S^{+(\rho+vt+b)}}[e^{\Li_{H}t}O_S]\|&\le 8Cr^dg\eta t+Cr^{-a}+2\eta.
\end{align}

For an interval of length $t\le T$, we can split it into $N\le tv/\ell+1$ short time intervals of lengths $s_i\le\ell/v$, which we order starting with $s_1$ from the end interval time at $t$.
Applying the triangle inequality and the preceding short time interval argument to each interval, and noting that the estimates with approximate (projected) evolved operators can only produce smaller norms since they are formed from unitary conjugations and condition expectations, gives an estimate like \eqref{eqn:short-error} for each short time step. 
More precisely, let $R_j:=\sum_{i=1}^j(\rho_i+vs_i+b)$, which describes the main observable spreading after backwards time $s_1+\cdots+s_j$ from the end time, and let $e^{\Li_Ht}=\oU_{N}(s_N)\cdots \oU_1(s_1)$ with $\oU_j(s_j)$ denoting the time-evolution corresponding to $s_j$, the $j$th interval counted backwards from the end time. Then letting $\Ex_j:=\Ex_{S^{+R_j}}$, and using $\Ex_N^2=\Ex_N$ and that the norm can only decrease under projection, we get
\begin{align*}
\|e^{\Li_Ht}O_S-\Ex_{S^{+R_N}}[e^{\Li_Ht}O_S]\|&\le\|e^{\Li_Ht}O_S-\Ex_N\oU_N\cdots\Ex_1\oU_1O_S\|+\|\Ex_N\oU_N\cdots\Ex_1\oU_1O_S-\Ex_N[e^{\Li_Ht}O_S]\|\\
&\le 2\|e^{\Li_Ht}O_S-\Ex_N\oU_N\cdots\Ex_1\oU_1O_S\|\\
&\le 2\sum_{j=1}^N\|(1-\Ex_j)\oU_j\Ex_{j-1}\oU_{j-1}\cdots \Ex_1\oU_1O_S\|, \numberthis\label{eqn:opN}
\end{align*}
using a telescoping sum for the last inequality. 
Each $\Ex_{j-1}\oU_{j-1}\cdots \Ex_1\oU_1O_S$ is automatically supported in $S^{+R_{j-1}}$ and has norm $\le\|O_S\|$.

The hypotheses $b\le\ell$, $v\le r^2$, $g\le r^2, T\le r$ with \eqref{eqn:short-error} then give total operator norm error \eqref{eqn:opN} at most
\begin{align}
2\sum_{i=1}^{tv/\ell+1}(8Cr^dg\eta s_i+Cr^{-a}+2\eta)\lesssim r^{d+3}\eta +(r^{-a}+\eta)r^3,
\end{align}
which gives the error bound $\eta'\lesssim r^{d+3}(\eta+r^{-a})$.
The total spreading distance $R_N$ is
\begin{align}
\sum_{i=1}^{tv/\ell+1}(\rho_i+ vs_i+b)&\le \sum_{i=1}^{tv/\ell+1}(C\ell gs_i+C_a\ell\log r+ vs_i+\ell)\lesssim_a (\ell g+v\log r)t+\ell\log r,
\end{align}
which gives the bounds $b'\lesssim_a\ell\log r$ and $v'\lesssim_a \ell g+v\log r$. 
\end{proof}

\section{Lower bounds for \texorpdfstring{$\alpha\le d$}{alpha≤d}}\label{sec:led-lower}

In this section, we prove the lower bounds for $\alpha\le d$ stated in Theorem~\ref{thm:led}(i). Recall when $\alpha\le d/2$, the bound becomes volume-dependent, and we work with a cube $Q_L$ of side length $L$.

\subsection{Truncation method}

To obtain a simple lower bound, which will be sufficient for the first three cases in the bound in Theorem~\ref{thm:led}(i), we use a standard truncation method to compare to a Hamiltonian with one site removed, see e.g. \cite[\S4.1]{chen2023speed}.
We delete all 2-site interaction terms involving a fixed site $x$, so a single-site observable $A_{x}$ at $x$ cannot evolve to other sites, then use Duhamel to compare the original Hamiltonian evolution to this non-spreading one for short times.

Let $H_0=\sum_{\substack{\{u,v\}\in\Lambda\\u,v\ne x}}h_{uv}(t)+\sum_u h_u(t)$ be the Hamiltonian with all 2-site terms involving $x$ removed, and let $A_x,B_y$ be single-site observables with $\|A_x\|,\|B_y\|\le1$, $x\ne y$. Then $e^{\Li_{H_0}t}A_x$ commutes with $B_y$, so $[e^{\Li_H t}A_x,B_y]=[e^{\Li_H t}A_x-e^{\Li_{H_0}t}A_x,B_y]$, and Duhamel with the bound $\|[D,B_y]\|_2\le 2\|D\|_2\|B_y\|$ gives
\begin{align*}
\|[e^{\Li_H t}A_x,B_y]\|_2&\le 4t\sup_{0\le s\le t}\|H(s)-H_0(s)\|_2.\numberthis
\end{align*}
Since $H(s)-H_0(s)=\sum_{u:u\ne x}h_{xu}(s)$, and $\Tr(h_{xu}^\dagger h_{xv})=0$ for $u\ne v$ by tracing over $u$ or $v$ first, we can directly estimate
\begin{align}
\|H(s)-H_0(s)\|_2^2&=\sum_{u\in Q_L:u\ne x}\|h_{xu}(s)\|_2^2\le \sum_{u\in Q_L:u\ne x}\frac1{|x-u|_\infty^{2\alpha}}\lesssim_{\alpha,d} \begin{cases}L^{d-2\alpha},&0\le\alpha<d/2\\
\log L,&\alpha=d/2\\
1,&\alpha>d/2
\end{cases}.
\end{align}
Thus for $A_x(t)=e^{\Li_Ht}A_x$, we obtain
\begin{align}
\|[A_x(t),B_y]\|_2&\lesssim t\begin{cases}L^{d/2-\alpha},&0\le\alpha<d/2\\
\sqrt{\log L},&\alpha=d/2\\
1,&\alpha>d/2
\end{cases}.
\end{align}
For the commutator Frobenius norm to reach $1/2$, we must have 
\begin{align}
t\gtrsim\begin{cases}
L^{\alpha-d/2},&0\le \alpha<d/2\\
1/\sqrt{\log L},&\alpha=d/2\\
1,&\alpha>d/2
\end{cases},
\end{align}
which gives the first three entries in the bound in Theorem~\ref{thm:led}(i). \qed

\subsection{Fractional-power logarithmic lower bounds}
We prove the last two entries in the bound in Theorem~\ref{thm:led}(i), the fractional-power log bounds for $3d/4<\alpha\le d$.
The proof bounds how much of the Pauli weight of $A_x(t)$ can appear at a site $y$ (i.e. the projection onto not the identity at $y$), related to some of the previous many-body quantum walk methods for bounding operator spreading \cite{chen2023speed,Lucas_2020_SYK,Faupin_2022_maximal, Faupin_2022_bh}.
Let $Q_y$ denote the orthogonal projection\footnote{One can check $Q_y^2=Q_y$, and $\Tr(B^\dagger(I_y\otimes\frac12\Tr_yA)=\frac12\Tr_{\Lambda\setminus\{y\}}(\Tr_yB^\dagger\Tr_yA)=\Tr((I_y\otimes\frac12\Tr_yB^\dagger)A)$, so $Q_y^*=Q_y$.} (in Frobenius/Hilbert--Schmidt inner product) onto terms with non-identity Pauli at site $y$, given by $Q_y(A)=A-I_y\otimes\frac12\Tr_y(A)$, for $\Tr_y$ the trace over $\C^2$ at site $y$. Then for $\|B_y\|\le1$,
\begin{align}\label{eqn:comm-log}
\|[A_x(t),B_y]\|_2=\|[Q_yA_x(t)+(1-Q_y)A_x(t),B_y]\|_2\le 2\|Q_yA_x(t)\|_2,
\end{align}
since $B_y=B_y\otimes I$ commutes with $(1-Q_y)A_x(t)=I_y\otimes\frac12\Tr_y(A_x(t))$ which is projected onto terms with the identity at $y$.

\begin{prop}[fractional-power log lower bound]\label{prop:lower-pl}
Let $3d/4<\alpha\le d$ and consider self-adjoint single-site operators $A_x,B_y$ with $\|A_x\|,\|B_y\|\le1$. Write $A=A_x=Q_yA+(1-Q_y)A$, and let $a_y(t):=\|Q_yA(t)\|_2^2$ and $R:=|x-y|_\infty$. Then as $R\to\infty$, we have $a_y(t)\to0$ for $t\lesssim(\log R)^{\alpha/d}$ if $3d/4<\alpha<d$, and for $t\lesssim \log R/\log\log R$ if $\alpha=d$. This implies $\|[A_x(t),B_y]\|_2\to0$ by \eqref{eqn:comm-log}, and thus proves the $3d/4<\alpha\le d$ lower bounds in Theorem~\ref{thm:led}(i).
\end{prop}

For this proof only, we take the convention $A(t):=U(t)AU(t)^\dagger$, where $\partial_t U(t)=-iH(t)U(t)$, so that $\dot A(t)=-i[H(t),A(t)]$. This will be equivalent to proving the result for the usual $A(t)=U(t)^\dagger AU(t)$ in Proposition~\ref{prop:lower-pl} by reversing the time, but this particular time indexing is more convenient so we don't have to index from the end time.
We start by differentiating $a_y(t):=\|Q_yA(t)\|_2^2$ in $t$ with this convention, using $\dot A(t)=-i[H(t),A(t)]$ and the Frobenius/Hilbert--Schmidt inner product $\langle A,B\rangle_\hs=2^{-|\Lambda|}\Tr_\Lambda(A^\dagger B)$ to obtain
\begin{align*}
\dot a_y(t)=2\Re\langle Q_y\dot A(t),Q_yA(t)\rangle_\hs&=-2\Re\langle i[H(t),A(t)],Q_yA(t)\rangle_\hs\\
&=-2\Re\langle i[H(t),(1-Q_y)A(t)],Q_yA(t)\rangle_\hs\\
&=-2\Re\langle i[\sum_{u\ne y}h_{uy},(1-Q_y)A(t)],Q_yA(t)\rangle_\hs ,\numberthis
\end{align*}
using that $\langle[H(t),Q_yA],Q_yA\rangle_\hs=0$ for $Q_yA$ self-adjoint, by cyclicity of the trace, to obtain the second line. For the third line, note that terms $h_u$ and $h_{uv}$ which do not involve $y$ commute with $Q_y$, so $[h_{uv},(1-Q_y)A(t)]=(1-Q_y)[h_{uv},A(t)]$ is orthogonal to $Q_yA(t)$. Also, the single-site term $h_y$ has $[h_y,(1-Q_y)A(t)]=0$.
Applying Cauchy--Schwarz then gives
\begin{align}\label{eqn:deriv-2norm}
\dot a_y(t)&\le 2a_y(t)^{1/2}\bigg\|\sum_{u:u\ne y}[h_{uy},(1-Q_y)A(t)]\bigg\|_2.
\end{align}

To bound the Frobenius norm in \eqref{eqn:deriv-2norm}, we cannot just use the triangle inequality since $\alpha\le d$ causes the sum over interaction strengths at any $y$ to diverge. To handle $u$ far away from $y$, we separate the sum over $u\ne y$ into regions $u\in S$ for $S$ a thin shell of radius $r$ and for a fixed Pauli operator direction, and use a concentration inequality to bound the Frobenius norm of the commutator. 
\begin{lem}[concentration bound]\label{lem:concentration}
Let $G=\sum_{i\in\Lambda}g_iP_i$ for real $g_i$ and a Pauli combination $P_i=a_iX_i+b_iY_i+c_iZ_i$ at site $i$, with real $a_i^2+b_i^2+c_i^2=1$.
Let $F$  be an operator which satisfies $FF^\dagger\le I$ and $F^\dagger F\le I$.
Then defining $\phi(x):=x\sqrt{\log(e/x)}$ for $0\le x\le1$, which is increasing, we have
\begin{align}
\|[G,F]\|_2^2&\lesssim \left(\sum_jg_j^2\right)^{1/2}\sum_i|g_i|\phi(\|Q_iF\|_2^2).
\end{align}
\end{lem}
\begin{proof}[Proof of Lemma~\ref{lem:concentration}]
Each Pauli combination $P_i$ has trace 0 and satisfies $P_i^2=I$, so has eigenvalues $\pm1$. For $z_i\in\{\pm1\}$ let $|z_i\rangle$ be a $z_i$ eigenstate of $P_i$. For $z\in\{\pm1\}^{\Lambda}$, let $|z\rangle$ denote the product state formed from the $|z_i\rangle$, and let $G(z):=\sum_i g_iz_i$ (the eigenvalues of $G$).
Then for $G$ eigenstates $|z\rangle,|w\rangle$ with $z,w\in\{\pm1\}^{\Lambda}$, we see $\langle z|[G,F]|w\rangle=(G(z)-G(w))\langle z|F|w\rangle$, and so
\begin{align}
\|[G,F]\|_2^2&=\sum_{z,w}2^{-|\Lambda|}|\langle z|F|w\rangle|^2|G(z)-G(w)|^2.
\end{align}
Letting $\pi(z,w):=2^{-|\Lambda|}|\langle z|F|w\rangle|^2$ be a (sub-probability) measure on $\{\pm1\}^{\Lambda^2}$, we see $\|[G,F]\|_2^2=\E_\pi|G(z)-G(w)|^2$ (note we use $\E_\pi$ for integration against $\pi$ even though $\pi$ need not be a probability measure). Since the row and column square sums of $F$ are $\le1$ by assumption, we see the marginals of $\pi$ are dominated by those of the uniform measure on $\{\pm1\}^{\Lambda}$.
For $z$ consisting of independent Rademacher variables (uniform $1/2$ probability for $\pm1$), calculating the moment generating function and using the exponential Markov inequality (or Hoeffding) combined with the above gives
\begin{align}
\pi\{|G(z)|\ge t\}\le \P_{\mathrm{unif}}[|G(z)|\ge t]&\le 2e^{-c't^2/\sum_j g_j^2}.
\end{align}
As a result, we can also bound
\begin{align}
\pi\{|G(z)-G(w)|\ge t\}&\le 4e^{-ct^2/\sum_j g_j^2},
\end{align}
since at least one of $|G(z)|$ or $|G(w)|$ must be $\ge t/2$ for the event.
Since $(G(z)-G(w))=\sum_ig_i(z_i-w_i)$, letting $E_i:=\{z_i\ne w_i\}$ and $\pi(E_i)=:\varepsilon_i$, we see
\begin{align*}
\|[G,F]\|_2^2=\E_\pi|G(z)-G(w)|^2&\le \E_\pi[|G(z)-G(w)|\sum_i2|g_i|\oneb_{z_i\ne w_i}]\\
&=2\sum_i|g_i|\int_{E_i}|G(z)-G(w)|\,d\pi\\
&=2\sum_i|g_i|\int_0^\infty \pi\{\oneb_{E_i}|G(z)-G(w)|>t\}\,dt\\
&\le 2\sum_i|g_i|\int_0^\infty\op{min}(\varepsilon_i,4e^{-ct^2/\sum_j g_j^2})\,dt\\
&\le C\sum_i|g_i|\left(\sum_jg_j^2\right)^{1/2}\varepsilon_i\sqrt{\log(e/\varepsilon_i)},
\numberthis\label{eqn:GF}
\end{align*}
the last line by splitting the integral where the two entries in the minimum cross.

Finally, we bound $\varepsilon_i=\pi(E_i)$ by noting that $\langle z|[P_i,F]|w\rangle=(z_i-w_i)\langle z|F|w\rangle$ and $|z_i-w_i|=2$ if $z_i\ne w_i$, so
\begin{align}
\varepsilon_i&=\frac14\|[P_i,F]\|_2^2\le \|Q_iF\|_2^2,
\end{align}
since $(1-Q_i)F$ projects onto the identity at site $i$ and so commutes with $P_i$.
As $\phi(x)=x\sqrt{\log(e/x)}$ is increasing on $[0,1]$, we obtain the lemma.
\end{proof}

Now that Lemma~\ref{lem:concentration} is proved, we return to \eqref{eqn:deriv-2norm}, in particular to bounding $\|\sum_{u:u\ne y}[h_{uy},(1-Q_y)A(t)]\|_2$.
If we tried to bound the sum over $u$ using a naive triangle inequality, we would obtain a bound $Cr^{d-\alpha}$ which does not decay for $\alpha\le d$. The bound is decent for small $r$ however, and for large $r$ we will instead use Lemma~\ref{lem:concentration}.

Let $F:=(1-Q_y)A(t)=I_y\otimes\frac12\Tr_yA$ be the projection of $A(t)$ onto terms with identity at $y$, and let $\rho\ge2$ be a cutoff for small vs large $r$ to be chosen later. 
Since $(1-Q_i)F$ acts as the identity on both $i$ and $y$, it commutes with $h_{iy}$, so $\|[h_{iy},F]\|_2=\|[h_{iy},Q_iF]\|_2\le 2\|h_{iy}\|\|Q_iF\|_2\le2\|h_{iy}\|\sqrt{a_i}$, using that $Q_i,Q_y$ commute.
For small $|i-y|_\infty<\rho$, we then have the simple bound
\begin{align}\label{eqn:c-near}
\sum_{\substack{i:i\ne y\\|i-y|_\infty<\rho}}\|[h_{iy},F]\|_2&\lesssim \sum_{\substack{i:i\ne y\\|i-y|_\infty<\rho}}r_{iy}^{-\alpha}\sqrt{a_i},\quad r_{iy}=|i-y|_\infty.
\end{align}

Next consider distant shells $S_k=\{i:\rho2^k\le|i-y|_\infty<\rho2^{k+1}\}$ for $k=0,1,2,\ldots$, and separate the interaction in each $S_k$ by Pauli $X,Y,Z$ directions at $y$,
\begin{align}
\mathfrak G_k:=\sum_{b=1}^3X_y^bG^b,\quad\text{for}\quad G^b:=\sum_{i\in S_k}\sum_{a=1}^3J_{iy}^{ab}X_i^a.
\end{align}
Then each $G^b$ is of the form $G^b=\sum_{i\in S_k}G_i^b$, with $G_i^b=\left(\sum_{a=1}^3(J_{iy}^{ab})^2\right)^{1/2}P_i$ for a rotated Pauli matrix $P_i$ of the form $a_iX_i+b_iY_i+c_iZ_i$ with $a_i^2+b_i^2+c_i^2=1$ as in Lemma~\ref{lem:concentration}.
Note that since $F$ acts as the identity on site $y$, the three terms $[X_y^bG^b,F]=X_y^b[G^b,F]$, $b=1,2,3$, are orthogonal in Frobenius/Hilbert--Schmidt inner product since distinct Paulis at site $y$ are orthogonal and $[G^b,F]$ acts as the identity at $y$.

Since $\|Q_iF\|_2=\|Q_i(1-Q_y)A(t)\|_2\le\|Q_iA(t)\|_2=\sqrt{a_i(t)}$ as $Q_i$ and $(1-Q_y)$ commute, applying Lemma~\ref{lem:concentration} once for each $b=1,2,3$ then gives for $r_k:=\rho2^k$,
\begin{align*}
\|[\mathfrak G_k,F]\|_2^2\le\sum_{b=1}^3\|[G^b,F]\|_2^2\lesssim \left(Cr_k^{d-2\alpha}\right)^{1/2}r_k^{-\alpha}\sum_{i\in S_k}\phi(a_i)&\lesssim r_k^{-(2\alpha-d/2)}\sum_{i\in S_k}\phi(a_i)\\
&\le\sqrt{\log(e/\delta)}r_k^{-(2\alpha-d/2)}\sum_{i\in S_k}(a_i+\delta),\numberthis\label{eqn:shell-estimate}
\end{align*}
using $\phi(x)\le\sqrt{\log(e/\delta)}(x+\delta)$ for any $0<\delta\le1$ (to be chosen later) in the last inequality.

Choose $\gamma$ strictly between $d$ and $2\alpha-d/2$, which is possible iff $\alpha>3d/4$.
Now we take a square root and sum over the shells; letting $\mathfrak G_{\mathrm{far}}:=\sum_{k\ge0}\mathfrak G_k$ and $A_k:=\sum_{i\in S_k}(a_i+\delta)$, then
\begin{align*}
\|[\mathfrak G_{\mathrm{far}},F]\|_2&\lesssim[\log(e/\delta)]^{1/4}\sum_{k\ge0}r_k^{-(\alpha-d/4-\gamma/2)}r^{-\gamma/2}\sqrt{A_k}\\
&\le [\log(e/\delta)]^{1/4}\left(\sum_{k\ge0}r_k^{-(2\alpha-d/2-\gamma)}\right)^{1/2}\left(\sum_{k\ge0}r_k^{-\gamma}A_k\right)^{1/2}\\
&\lesssim [\log(e/\delta)]^{1/4}\rho^{-(\alpha-d/4-\gamma/2)}\left(\sum_{i:r_{iy}\ge\rho}r_{iy}^{-\gamma}(a_i+\delta)\right)^{1/2},\quad\text{for } r_{iy}=|i-y|_\infty.\numberthis
\end{align*}

Returning to the bound for $\dot a_y(t)$ in \eqref{eqn:deriv-2norm}, and combining with the near bound \eqref{eqn:c-near} and $\sqrt{a}\sqrt{b}\le\frac12(a+b)$, we obtain
\begin{align*}
\dot a_y(t)&\lesssim \sum_{r_{iy}<\rho}r_{iy}^{-\alpha}(a_i+a_y)+[\log(e/\delta)]^{1/4}\rho^{-(\alpha-d/4-\gamma/2)}\left(\rho^{d-\gamma}\left[\left(\rho^{\gamma-d}\sum_{r_{iy}\ge\rho}r_{iy}^{-\gamma}a_i\right)+\delta\right]\right)^{1/2}\sqrt{a_y}\\
&\lesssim a_y\begin{cases}\rho^{d-\alpha},&\alpha<d\\
(\log\rho),&\alpha=d\end{cases}+\sum_{i:0<r_{iy}<\rho} r_{iy}^{-\alpha}a_i+ [\log(e/\delta)]^{1/4}\rho^{3d/4-\alpha}\left(a_y+\rho^{\gamma-d}\sum_{r_{iy}\ge\rho}r_{iy}^{-\gamma}a_i+\delta\right).\numberthis\label{eqn:deriv-bound}
\end{align*}
Equation~\eqref{eqn:deriv-bound} can be written in a simpler form in terms of convolution since $r_{iy}=|i-y|_\infty$. For $i\in\Z^d$, define
\begin{align*}
w_\rho(i):=\begin{cases}
|i|_\infty^{-\alpha},&0<|i|_\infty<\rho\\
[\log(e/\delta)]^{1/4}\rho^{3d/4-\alpha}\rho^{\gamma-d}|i|_\infty^{-\gamma},&|i|_\infty\ge\rho\\
0,&i=0
\end{cases}.
\end{align*}
To make these weights into probabilities so we can easily use probability tail bounds later, we choose a large normalization constant $M$ via, for $\alpha<d$,
\begin{align*}
\sum_{i\in\Z^d} w_\rho(i)&\le C\rho^{d-\alpha}+[\log(e/\delta)]^{1/4}\rho^{3d/4-\alpha+\gamma-d}C\rho^{d-\gamma}
\le C_1(\rho^{d-\alpha}+[\log(e/\delta)]^{1/4}\rho^{3d/4-\alpha})=:M.
\end{align*}
For $\alpha=d$, the $\rho^{d-\alpha}$ term in $M$ is instead $\log\rho$.
Define $p(z)=\frac1Mw_\rho(z)$ for $z\ne0$, and $p(0)=1-\frac1M\sum_{z\ne0}w_\rho(z)$, so that $p$ is a probability measure on $\Z^d$.
For convolution $(p*a)_y=\sum_{i\in\Z^d}p(y-i)a_{i}=\sum_{i\in\Z^d}p(i)a_{y-i}$ (extend $a$ by zero outside $\Lambda$ if needed), we can bound \eqref{eqn:deriv-bound} as
\begin{align}\label{eqn:de-inequality}
\dot a_y(t)&\le CM\left[a_y(t)+(p*a)_y(t)+\delta\right].
\end{align}

We next ``solve'' this differential inequality. 
Let $\oneb$ denote the constant function on the lattice, $Pf:=p*f$, and $s\le t$, and use an integrating factor, e.g. note that
\begin{align*}
\frac{d}{ds}[e^{CM(t-s)(I+P)}a(s)]&=e^{CM(t-s)(I+P)}[-CM(I+P)a(s)+\dot a(s)]\\
&\le e^{CM(t-s)(I+P)}CM\delta\oneb,\numberthis
\end{align*}
using \eqref{eqn:de-inequality} and that $e^{u(I+P)}=e^u\sum_{n=0}^\infty\frac{u^n}{n!}P^n$ preserves pointwise inequalities of functions, $f_y\le g_y$, since $u\ge0$ and $P$ preserves this. Integrating from $0$ to $t$ and using $a(0)\le\oneb_{\{x\}}$ then gives
\begin{align*}
a_y(t)&\le \left[e^{CMt(I+P)}a(0)+CM\delta\int_0^t e^{CM(t-s)(I+P)}\oneb\,ds\right]_y\\
&\le e^{CMt}\sum_{n=0}^\infty\frac{(CMt)^n}{n!}p^{*n}(y-x)+\frac{\delta}{2}(e^{2CMt}-1). \numberthis\label{eqn:a-de}
\end{align*}

To estimate $p^{*n}(y-x)$, let $Z_1,\ldots,Z_n$ be i.i.d. $\Z^d$-valued random variables with common distribution $p$, so that $p^{*n}(y-x)=\P[Z_1+\cdots+Z_n=y-x]$. We will bound this probability using Markov's inequality, so we want to estimate moments of $p$. Letting $0<q<\op{min}\{1,\gamma-d\}$, we can quickly estimate (the following in $\ell^\infty$ norm)
\begin{align*}
\E|Z_i|^q&=M^{-1}\sum_{0<|z|<\rho}|z|^q|z|^{-\alpha}+M^{-1}\sum_{|z|\ge\rho}|z|^q[\log(e/\delta)]^{1/4}\rho^{3d/4-\alpha}\rho^{\gamma-d}|z|^{-\gamma}\\
&\lesssim M^{-1}\left[\rho^{d+q-\alpha}+[\log(e/\delta)]^{1/4}\rho^{3d/4-\alpha}\rho^{q}\right]
\lesssim \rho^q,\numberthis\label{eqn:Z-moment}
\end{align*}
using $\alpha\ge 3d/4$.
Then applying Markov's inequality as well as $\E|Z_1+\cdots+Z_n|^q\le\E|Z_1|^q+\cdots+\E|Z_n|^q$ since $q\le1$ gives for $R=|y-x|_\infty$,
\begin{align}
p^{*n}(y-x)\le \P[|Z_1+\cdots+Z_n|\ge R]\le  Cn\left(\frac{\rho}{R}\right)^q.
\end{align}
Finally, from \eqref{eqn:a-de} we obtain for $y\ne x$,
\begin{align}\label{eqn:ay-final}
a_y(t)&\lesssim CMte^{2CMt}\left(\frac{\rho}{R}\right)^q+\frac\delta2(e^{2CMt}-1).
\end{align}

To complete the proof, we choose $\delta$ and $\rho$. For $R=|x-y|_\infty$ sufficiently large, set $\delta:=R^{-\gamma}$ and $\rho:=\lceil(\log R)^{1/d}\rceil$.
For $3d/4<\alpha<d$, let $t\le c_1(\log R)^{\alpha/d}$ with $c_1$ small enough so that $2CMt\le\frac{q}{2}\log R$ so $e^{2CMt}\le R^{q/2}$.
Then for such $t$, \eqref{eqn:ay-final} gives
\begin{align}
a_y(t)&\lesssim (\log R)^{1+q/d}R^{-q/2}+R^{-\gamma+q/2}\to0,\quad\text{as $R\to\infty$},
\end{align}
since $q>0$, $\gamma>d\ge1>q/2$, and $M=O(\rho^{d-\alpha})=O((\log R)^{1-\alpha/d})$. 
If $\alpha=d$, then we instead have $M=C_1(\log\rho+[\log(e/\delta)]^{1/4}\rho^{-d/4})=O(\log\rho)=O(\log\log R)$. Then for $t\le c_1\log R/\log\log R$, \eqref{eqn:ay-final} gives $a_y(t)\to0$ as $R\to\infty$.
This proves Proposition~\ref{prop:lower-pl}. \qed

\section{Protocols for \texorpdfstring{$0\le\alpha<d$}{0≤alpha≤d}}\label{sec:protocol}

In this section, we prove Theorem~\ref{thm:led}(ii).

\subsection{Protocol for \texorpdfstring{$0\le\alpha\le d/2$}{0≤alpha≤d/2}}\label{subsec:decay-protocol}
We start by proving the first bound $T\lesssim(\log L)L^{\alpha-d/2}$ in \eqref{eqn:alphad-protocol} in Theorem~\ref{thm:led}(ii).
Let $Q_L\subset\Z^d$ be a cube with side-length $L$. We show there is a protocol, i.e.~a Hamiltonian \eqref{eqn:H}, which takes time $O((\log L)L^{\alpha-d/2})$ and produces large enough Frobenius norm commutator $\|[A(t),Z_y]\|_2$ starting from $A=X_x$. 
The protocol is a simple adaptation of the random $\alpha>d$ protocols in \cite{chen2021optimal}, with some technical simplifications. In particular, we use the same random Clifford layers (or depolarizing channel), alternated with a similar random $ZZ$ Hamiltonian pulse, though without any recursive procedure.
Since we primarily consider the strongly long-range regime $\alpha\le d/2$, it will turn out to be sufficient to take the all-to-all Hamiltonian
\begin{align}\label{eqn:H-zz}
H_\sigma^{(Q_L)}:=\sum_{\{i,j\}\subset Q_L}\sigma_{ij}Z_iZ_j,
\end{align}
for independent uniform random signs $\sigma_{xy}\in\{\pm1\}$. 
We will think of normalizing the Hamiltonian by $L^{-\alpha}$ so that it meets the $r^{-\alpha}$ power-law limit.
(The protocol works for any $\alpha$, but only gives a useful time for $\alpha\le d/2$; we will construct polylog protocols for $d/2<\alpha<d$ in Section~\ref{subsec:polylog-protocol}.) Since the different terms $Z_iZ_j$ commute, we can study the Hamiltonian's effect by considering one edge $\{i,j\}$ at a time.

The all-to-all interactions give permutation invariance which will be used to average over the target site $y$.
For the other operation, a random Clifford gate acts by conjugation on Paulis $X,Y,Z$ to output one of $\{\pm X,\pm Y,\pm Z\}$ each with equal probability. The random Clifford layer applies an independent random Clifford gate at each site. We will alternate random Clifford layers and $ZZ$ Ising pulses in the protocol, and analyze how the support of $A(t)$ spreads under the protocol. 

\begin{figure}[htb]
\begin{tikzpicture}[scale=.7]
\begin{scope}
\draw[gray] (-2,-2) grid (2,2);
\fill[blue] (0,0) circle (1mm) node[below left] {$X$};
\draw[->] (2.5,0)--++(1,0);
\node[above] at (3,0) {$e^{i\theta H}$};
\end{scope}

\begin{scope}[xshift=6cm]
\draw[gray] (-2,-2) grid (2,2);
\fill[blue] (0,0) circle (1mm) node[below left] {$Y$};
\fill[black] (0,-1) circle (1mm) node[below left] {$Z$};
\fill[black] (2,2) circle (1mm) node[below left] {$Z$};
\fill[black] (-1,2) circle (1mm) node[below left] {$Z$};
\draw[->] (2.5,0)--++(1,0);
\node[above] at (3,0) {Clifford};
\end{scope}

\begin{scope}[xshift=12cm]
\draw[gray] (-2,-2) grid (2,2);
\fill[blue] (0,0) circle (1mm) node[below left] {$Z$};
\fill[black] (0,-1) circle (1mm) node[below left] {$X$};
\fill[black] (2,2) circle (1mm) node[below left] {$Z$};
\fill[black] (-1,2) circle (1mm) node[below left] {$Y$};
\end{scope}
\end{tikzpicture}
\caption{Illustration of one step of the protocol in Theorem~\ref{thm:led}(ii). Under averaging over the random signs $\sigma_{ij}$ in the $ZZ$ Hamiltonian, the Pauli string probabilities evolve according to a random classical process shown in the grid above. Starting from the Pauli string $X$ supported at site $x$, we want to grow the support to macroscopic size. The $ZZ$ Hamiltonian pulse has a chance of turning on any site from $I$ to $Z$, or turning off a site from $Z$ to $I$, as well as flipping $X\leftrightarrow Y$. The random Clifford layer permutes $X,Y,Z$ on sites inbetween $ZZ$ pulses. We show repeating this process grows the support rapidly, which will be used to show the expected value of the Frobenius commutator over the protocol, $\E\|[A_x(t),Z_y]\|_2^2$, is large.}\label{fig:protocol1}
\end{figure}
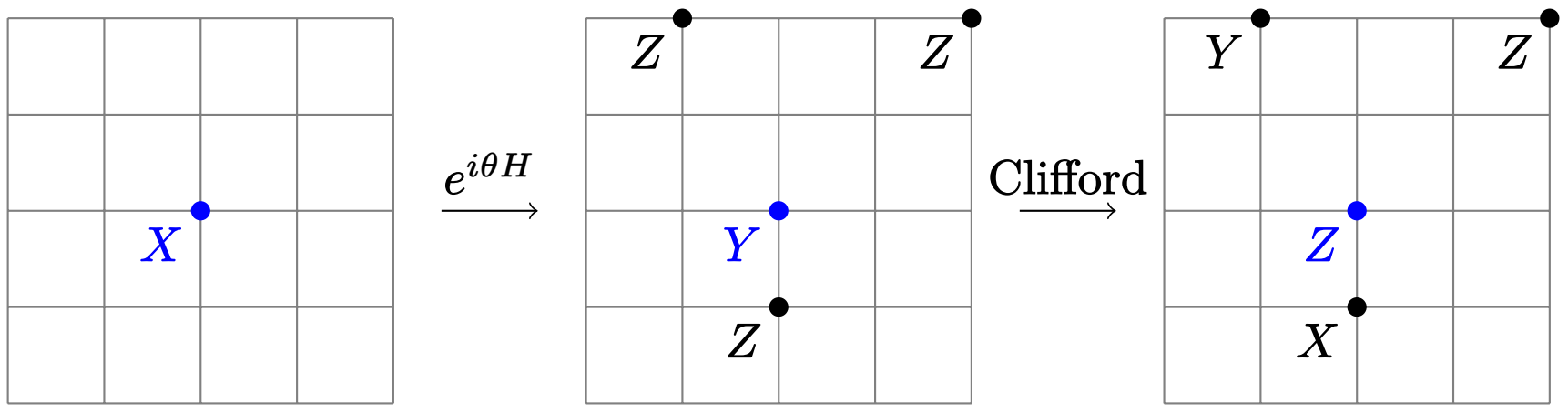

As in \cite{chen2021optimal}, the protocol is associated with a classical probabilistic problem.
Under averaging over the random signs $\sigma_{ij}$, a small duration $\theta$ pulse of $H$ evolves Pauli strings $P\in\{X,Y,Z,I\}^{\otimes Q_L}$ according to a classical process which flips sites $I\leftrightarrow Z$, and $X\leftrightarrow Y$, with a small probability depending on $\theta$ (Figure~\ref{fig:protocol1}). A random Clifford layer then randomly permutes $X,Y,Z$ at each site (leaving $I$ alone).
Starting from $X_x$ at site $x$, we are interested in how this process grows the support (non-$I$ sites) of the Pauli string. The $ZZ$ pulse turns on some random sites from $I$ to $Z$, and also sends some $Z$ back to $I$. It can flip $X\leftrightarrow Y$, but this doesn't affect the support. The random Clifford layer turns some of $Z$s to $X$ or $Y$, which the $ZZ$ pulse then cannot send back to $I$. This layer also means we don't need to keep precise track of which Pauli is at each site since they are  randomized each round.

Let $P$ be a Pauli string, i.e. tensor product of $X,Y,Z,I$ operators at each site. Distinct Pauli strings are orthonormal in normalized Hilbert-Schmidt norm.
We first determine the result of applying the random $ZZ$ Ising Hamiltonian \eqref{eqn:H-zz} to a Pauli string.

\begin{lem}[$ZZ$ Ising pulse]\label{lem:zz}
Let $P$ be a Pauli string, and apply a duration $\theta$ pulse by \eqref{eqn:H-zz}. The resulting output is a superposition $\sum_QC_QQ$ of Pauli strings $Q$, with probabilities $\E_\sigma|C_Q|^2$ determined by the following classical process:
Call a site active if its Pauli is $X$ or $Y$, and inactive if its Pauli is $I$ or $Z$.
For each edge $\{i,j\}$, do nothing if both endpoints are active or both are inactive. Otherwise, if exactly one endpoint is active: do nothing to the current Pauli string $P$ with probability $\cos^2(2\theta)$, and flip to $Z_iZ_jP$ with probability $\sin^2(2\theta)$, indepedently for each edge. The expected weight $\E_\sigma|C_Q|^2$ is exactly the probability of obtaining $Q$ through this classical process.

The number of active sites, $a$, does not change during the $ZZ$ pulse, and the probability for a particular inactive site to flip $I\leftrightarrow Z$ in the classical process is
\begin{align}
p_a&:=\P[I\leftrightarrow Z]=\frac{1-(\cos 4\theta)^a}{2}.
\end{align}
\end{lem}
\begin{proof}[Proof of Lemma~\ref{lem:zz}]

A single $Z_iZ_j$ evolves $P$ by unitary conjugation, $e^{i\theta\sigma_{ij}Z_iZ_j}Pe^{-i\theta\sigma_{ij}Z_iZ_j}$, and either commutes or anticommutes with a Pauli string $P$. In the commuting case, the evolution doesn't change $P$. Otherwise, if they anticommute, then power series expansion shows
\begin{align}\label{eqn:cos-sin}
e^{i\theta\sigma_{ij}Z_iZ_j}Pe^{-i\theta\sigma_{ij}Z_iZ_j}&=\cos(2\theta)P+i\sigma_{ij}\sin(2\theta)Z_iZ_jP,
\end{align}
which means the evolution is a superposition of $P$ and the new Pauli string $Z_iZ_jP$. Since $\sigma_{ij}\in\{\pm1\}$ and distinct Pauli strings are orthogonal, the probability amplitude is $\cos^2(2\theta)$ for $P$, and $\sin^2(2\theta)$ for $Z_iZ_jP$.
Recall a site is active if its Pauli is $X$ or $Y$, and inactive if its Pauli is $I$ or $Z$.
Since Pauli matrices anticommute, the Pauli string $P$ anticommutes with $Z_iZ_j$ iff exactly one of the endpoints of $\{i,j\}$ is active. Also, if $P$ anticommutes with $Z_iZ_j$, then so does $Z_iZ_jP$.
Then for a superposition $\sum_P C_PP$ of Pauli strings, evolution by $\sigma_{ij}Z_iZ_j$ gives
\begin{align*}
&e^{i\theta\sigma_{ij}Z_iZ_j}\sum_P C_PP e^{-i\theta\sigma_{ij}Z_iZ_j}\\
&=\sum_P \begin{cases}C_P(\cos(2\theta)P+i\sigma_{ij}\sin(2\theta)Z_iZ_jP),&\text{exactly 1 endpoint of $\{i,j\}$ active in $P$}\\
C_PP,&\text{otherwise}
\end{cases}\\
&=\sum_P \begin{cases}(C_P\cos(2\theta)+\zeta_{ij,P}\sigma_{ij}\sin(2\theta)C_{Z_iZ_jP}) P,&\text{exactly 1 endpoint of $\{i,j\}$ active in $P$}\\
C_PP,&\text{otherwise}
\end{cases}=:\sum_PC_P'P,\numberthis
\end{align*}
for a phase $\zeta_{ij,P}$. Because the signs $\sigma_{ij}$ are independent random with mean zero (and independent of any previous steps in the process), we see $\E_{\sigma}[\sigma_{ij}C_P\overline{C_Q}]=0$, which implies that conditioned on the $C_P$,
\begin{align}\label{eqn:new-C}
\E_{\sigma}|C_P'|^2&=\begin{cases}\cos^2(2\theta)|C_P|^2+\sin^2(2\theta)|C_{Z_iZ_jP}|^2,&\text{exactly 1 endpoint of $\{i,j\}$ active in $P$}\\
|C_P|^2,&\text{otherwise}
\end{cases}.
\end{align}
Equation~\eqref{eqn:new-C} shows that under evolution by the full all-to-all Hamiltonian \eqref{eqn:H-zz}, whose individual summands commute, the expected Pauli transition probabilities from applying an edge $\{i,j\}$ are unchanged if both endpoints are active or both inactive in the current $P$, and $\cos^2(2\theta)$ to stay at $P$ and $\sin^2(2\theta)$ to flip to $Z_iZ_jP$ if exactly 1 endpoint is active in the current $P$. This proves the first part of the lemma.

The above then implies the classical process probability to flip $I\leftrightarrow Z$ at a single site.
Under the all-the-all connectivity of \eqref{eqn:H-zz} and the above classical process, each nonactive ($Z$ or $I$) site has the same number $a$ of neighboring active sites. It flips if there are an odd number of flips out of the $a$ active edge interactions. All the $\sigma_{ij}$ are independent, so the expected probability that $I$ flips to $Z$ at any single site is thus
\begin{align*}
p_a&=\P[I\leftrightarrow Z]= \P[\op{Bin}(a,\sin^2(2\theta))\text{ is odd}]\\
&=\sum_{\substack{k=0\\k\text{ odd}}}^a\binom{a}{k}\sin^{2k}(2\theta)(1-\sin^2(2\theta))^{a-k}
=\frac{1-(\cos 4\theta)^a}{2},\numberthis
\end{align*}
where for the last equality we expanded $(x+y)^a-(x-y)^a$.
\end{proof}

\begin{proof}[Proof of the first bound in Theorem~\ref{thm:led}(ii)]
The protocol repeats the following two steps.
We first apply a random Clifford gate layer, using the allowed unbounded single-site terms to implement each gate using a single-site Hamiltonian in arbitrarily fast time. Then we apply the random Ising Hamiltonian pulse, and estimate how the Pauli string $P$ has spread. Our goal is to show that the support of $P$ reaches size $N/4$ quickly, which will then imply a large enough expected commutator overlap with any fixed $y$ by permutation invariance of the non-$x$ target sites under the protocol.

For an initial operator $A$, we will consider the time-evolved Frobenius norm commutator with $Z_y$, averaged over the random protocol, which we write as $\E\|[A(t),Z_y]\|_2^2$. Writing $A(t)=\sum_P C_PP$, then $[A(t),Z_y]=\sum_PC_P[P,Z_y]=\sum_{P:P_y=X\text{ or }Y}2C_PPZ_y$, and so
\begin{align}\label{eqn:commutator-exp}
\E\|[A(t),Z_y]\|_2^2=4\sum_{P:P_y=X\text{ or }Y}\E|C_P|^2.
\end{align}
We are thus interested in $\E|C_P|^2$, which is described by the classical spreading process described in Lemma~\ref{lem:zz}.

We describe the role of the random Clifford layer in the classical process.
When applying a random 1-site Clifford gate to each site, each site in the support of a Pauli string $P$ (where it is not $I$) becomes $X,Y,Z$ (up to a random sign) with equal probability.
Recall an active site is one whose Pauli is $X$ or $Y$.
After applying the random Clifford layer, the number of active sites in the classical process is distributed as $a\sim\op{Bin}(|\op{supp}P|,2/3)$. 
Conditioned on $a$, by Lemma~\ref{lem:zz} the size of the support in the classical process after applying the $ZZ$ Hamiltonian is
\begin{align}\label{eqn:newsupport}
s'\sim a+\op{Bin}(s-a,1-p_a)+\op{Bin}(N-s,p_a),
\end{align}
where $s$ is the size of the previous support, $N$ is the number of sites in $Q_L$, and the two binomial distributions (coming from not flipping $Z$ sites, and flipping $I$ sites) are independent.

For the protocol, we start from $A=X_x$ at site $x$.
\begin{enumerate}[leftmargin=*]
\item We first show that a single $ZZ$ pulse increases the Pauli support of the classical process to a large fixed value $m$ (to be chosen later) with high probability. (We skip the random Clifford layer for this first step.) The site $x$ is active and $s=a=1$, so the new support after the $ZZ$ pulse by Lemma~\ref{lem:zz} and \eqref{eqn:newsupport} is
\begin{align}
s_1=1+B,\quad B\sim \op{Bin}(N-1,p_1),\quad p_1=\frac{1-\cos(4\theta)}{2}=\sin^2(2\theta).
\end{align}
The expected number of new occupied (not $I$) sites is $(N-1)p_1=(N-1)\sin^2(2\theta)\approx 4N\theta^2$ for small $\theta$ and large $N$. We want this to be a large constant, so we will take the pulse time $\theta=C_1N^{-1/2}$ for a constant $C_1\ge\sqrt{m}$ to be determined. Then by the Chernoff bound,
\begin{align}\label{eqn:fail-m}
\P[s_1<m]&\le Ce^{-cm}.
\end{align}

\item Next, we show that subsequent random Clifford plus $ZZ$ pulses double the support $s$ with high probability, for $m\le s\le \rho N$, with $\rho$ a small fixed constant to be determined. If $s$ is the current support, a random Clifford layer gives the number of active sites as $a\sim\op{Bin}(s,2/3)$, and so $\P[a>s/3]\ge 1-e^{-cs}$. If $a\ge s/3$, then since $\cos x\le e^{-x^2/2}$ for small $x$, Lemma~\ref{lem:zz} gives
\begin{align}\label{eqn:a-bound}
p_a=\frac{1-(\cos(4C_1N^{-1/2}))^a}{2}&\ge \frac{1-e^{-8C_1^2a/N}}{2}\ge \frac{1-e^{-8C_1^2s/(3N)}}{2}.
\end{align}
Take $C_1$ large enough (still to be fully determined later) so that $\frac12(1-e^{-8C_1^2\rho/3})\ge1/4$; then $p_a\ge s/(4\rho N)$ for $s\le\rho N$ by concavity of $1-e^{-x}$.
From \eqref{eqn:newsupport}, let $B\sim\op{Bin}(N-s,p_a)$ be the new occupied sites from flipping $I$ to $Z$, so that conditioned on $a\ge s/3$, $\E B\ge\frac{1-\rho}{4\rho}s$. Take e.g. $\rho=1/32$ so that $\E B\ge 4s$. Then since $s'\ge B$, Chernoff again gives
\begin{align}\label{eqn:doubling-failure}
\P[s'<2s|s] &\le \P[B<2s|a\ge s/3]+\P[a<s/3]\le 2e^{-cs}.
\end{align}

Now we repeat this process until we reach $\rho N$ occupied sites. If we have $j$ successful doublings, the support has size at least $2^jm$. The probability that the doubling procedure failed at any stage, using \eqref{eqn:doubling-failure}, is at most
\begin{align}
\P[\text{a doubling fails before reaching $\rho N$}|s_1\ge m] &\le \sum_{j=0}^\infty 2e^{-2^jcm}\le \frac{2e^{-cm}}{1-e^{-cm}}.\label{eqn:fail-doubling}
\end{align}
Choose $m$ so this is $\le1/100$. Under successful doublings, reaching $\rho N$ requires less than e.g. $K:=\log_2\frac{N}{4m}+10=O(\log N)$ iterations.

\item Finally, we show that once the evolution has large support, it remains with large support. This lets us have a fixed non-random end time over realizations of the protocol.
Suppose the support has reached $s\ge \rho N$. We will show the next round produces support at least $N/4$ with probability at least $1-e^{-cN}$. The random Clifford layer gives the active site bound $\P[a<\rho N/3]\le e^{-c\rho N}$. If $a\ge \rho N/3$, then use \eqref{eqn:a-bound} and finally choose $C_1$ large enough for all previous steps, and now also so that $p_a\ge \frac12(1-e^{-8C_1^2\rho/3})\ge 3/8$. Using \eqref{eqn:newsupport}, with $p_a\ge 3/8$ and also noting that $p_a\le1/2$ so $1-p_a\ge1/2$, shows that $s'$ stochastically dominates $\op{Bin}(N,3/8)$. Thus, the Chernoff bound again gives
\begin{align}
\P[s'< N/4|s\ge \rho N] &\le e^{-c N}+\P[a<\rho N/3]\le e^{-cN}.
\end{align}
Recalling we reach support size $\rho N$ before $K=\log_2N+10$, we bound the probability of failure between reaching $\rho N$ and reaching $N/4$ at most $Ke^{-cN}=o(1)$. Thus combining the above, \eqref{eqn:fail-m}, and \eqref{eqn:fail-doubling}, we obtain probability of failure at time $K$ at most $e^{-cm}+\frac{1}{100}+o(1)$. Taking appropriate $m,C_1$ gives for sufficiently large $N$, e.g.
\begin{align}\label{eqn:support}
\P[s_K\ge N/4]&\ge 3/5,\quad\Longrightarrow\quad \E s_K\ge 3N/20.
\end{align}

Since we use $O(\log N)$ rounds, with pulse times $\theta=O(N^{-1/2})$ each round, the total time for the protocol with adding an $L^{-\alpha}$ normalization in \eqref{eqn:H-zz} is
\begin{align}
T\lesssim (\log N)N^{-1/2}L^\alpha\sim_d (\log L)L^{\alpha-d/2}.
\end{align}
\end{enumerate}

Finally, apply one last independent random Clifford layer to randomize any Pauli at site $y$. Then from \eqref{eqn:commutator-exp},
\begin{align}
\E\|[A(t),Z_y]\|_2^2&=\frac83\sum_{P:y\text{ occupied}}\E|C_P|^2=\frac83\P[y\text{ occupied}],
\end{align}
where we call a site occupied if it is in the support of $P$, i.e. $P_y$ is $X,Y$, or $Z$, not the identity. 
Since the $ZZ$ Hamiltonian and random Clifford gates and permutation symmetric over the sites, occupation for all non-$x$ sites are equivalent. So using \eqref{eqn:support}, we obtain for sufficiently large $N$
\begin{align}
\P[y\text{ occupied}]=\P[y\in\supp P]&\ge \frac{\E s_K-1}{N-1}\ge \frac{3N/20-1}{N-1}\ge \frac18,
\end{align}
which implies
\begin{align}
\E\|[A(t),Z_y]\|_2^2&=\frac83\P[y\text{ occupied}]\ge\frac13\ge\frac14.
\end{align}
Therefore, there is some deterministic realization with Frobenius commutator norm at least $1/2$.
\end{proof}

\subsection{Polylogarithmic Protocol for \texorpdfstring{$d/2< \alpha <d$}{d/2<alpha<d}}\label{subsec:polylog-protocol}
We can modify the protocol in the prior section and \cite{chen2021optimal} slightly to make it give a polylogarithmic run time when $\alpha \in (d/2,d)$ as well. The key idea is to recursively spread an operator through a hierarchy of increasingly large hypercubes, guaranteeing, with sufficiently high probability, that at each scale the operator has spread sufficiently densely throughout each hypercube. Let $\ell_0,m_1,\cdots,m_k\in \mathbb{N},$ $m_j\geq 2$ for all $j,$ be such that $R:=\vert x-y\vert_\infty=\ell_0\prod_{j=1}^k m_j.$ Let $\ell_j := \ell_0 \prod_{l=1}^jm_l.$ We then let $\mathcal{Q}_j$ be nested partitions of the hypercube $Q_R$ of side-length $R$ into hypercubes of side length $\ell_j.$
Let
\begin{align}
    U_{j+1}:= U_{j}\exp\left(-i\pi\frac{\ell_0^\alpha}{8}\sum_{q\in \mathcal{Q}_{0}}\ell_0^{-\alpha}H_{\sigma}^{(q)}\right) \mathcal{C} \exp\left(-it_j\sum_{q\in \mathcal{Q}_{j+1}}\ell_{j+1}^{-\alpha}H_{\sigma'}^{(q)}\right)\mathcal{C}'U_{j}',\label{eq:recursion}
\end{align} where $t_j:= \ell_{j+1}^\alpha\min\left\{\frac{\ell_0^{d/2}}{\ell_j^d},\frac{\pi}{8}\right\},$ $U_0:=I,$ and the priming indicates that the underlying randomness is independent of that used in the other symbols. Our protocol consists of $U_k$ followed by some final post-processing we will elaborate on later to provide sufficient likelihood of the target site being active. We begin by establishing the relevant property of $U_j.$
\begin{prop}
    Let $\varepsilon_j:=\left[2^j\left(\frac{\ell_j}{\ell_0}\right)^d-1\right]e^{-\frac{\ell_0^d}{12}}.$ Then, for any $Q\in\mathcal{Q}_j$ and Pauli-string $P$ with at least one active site in $Q,$ if we let $P'$ be a Pauli string sampled from $U_j^\dagger PU_j$ with probability equal to the Pauli string's squared coefficient, then
    \begin{align}
        \mathbf{P}[\forall D\in\mathcal{Q}_0,\ D\subseteq Q\colon P'\text{ has active site in D}]\geq 1-\varepsilon_j.\label{eq:inductcond}
    \end{align}
\end{prop}
\begin{proof}
    We prove this by induction in $j$. We begin by noting that for $j=0,$ $U_j=I,$ meaning the condition in Eq.~(\ref{eq:inductcond}) is satisfied for $Q,$ and, as all other elements of $\mathcal{Q}_0$ are disjoint from $Q,$ the assumption that $P$ has an active site in $Q$ trivially satisfies Eq.~(\ref{eq:inductcond}). Thus, the base case is satisfied.

    We now establish the inductive step. Suppose Eq.~(\ref{eq:inductcond}) is true for $U_j$. Consider some $Q\in\mathcal{Q}_{j+1},$ and a Pauli string $P$ with at least one active site in a $Q_*\in \mathcal{Q}_j$ that is inside $Q.$ Eq.~(\ref{eq:inductcond}) implies that, after the application of the leftmost $U_j$ in Eq.~(\ref{eq:recursion}), there are active sites in all $D\subseteq Q_*,$ $D\in\mathcal{Q}_0$ with a probability $\geq1-\varepsilon_j$. By Lemma~\ref{lem:zz}, conditioned on some Pauli string with active sites in all $D\subseteq Q_*,$ $D\in \mathcal{Q}_0,$, after the application of the normalized $H_\sigma^{(q)}$s for all $q\in\mathcal{Q}_0,$ and the application of $\mathcal{C}$, the event of each site being active is independent and has a probability of at least $1/3$. We thus have that, by an application of Hoeffding's inequality,
    \begin{align}\mathbf{P}[\text{num. sites active in }Q_*< \vert Q_*\vert/8] \leq \varepsilon_j + e^{-\frac{\vert Q_*\vert}{12}}\label{eq:Qover}.\end{align}

    We next condition on some Pauli string for which the number of active sites in $Q_*$ is $a\geq \vert Q_*\vert/8.$ After the application of all the normalized $H_{\sigma'}^{(q)}$s for $q\in\mathcal{Q}_{j+1}$ and the next layer of cliffords $\mathcal{C}',$ by Lemma~\ref{lem:zz} we see that, independently, for all $x\in Q,$ 
    \begin{align}
        \mathbf{P}\left[x\text{ active}\right]&\geq \frac{1-\cos^a\frac{4t_j}{\ell_{j+1}^\alpha}}{3}\nonumber\\
        &\geq \frac{1}{6}\left(\frac{\ell_0}{\ell_{j}}\right)^d.
    \end{align} Thus, for any $q\in \mathcal{Q}_j$ satisfying $q\subseteq Q,$ the probability that no site in $q$ is active is no larger than
    \begin{align}
        \left[1-\frac{1}{6}\left(\frac{\ell_0}{\ell_{j}}\right)^d\right]^{\ell_j^d} \leq e^{-\ell_0^d/6}.
    \end{align}
    By assumption, for any $q\in \mathcal{Q}_j$ satisfying $q\subseteq Q,$ we see that after applying $U_j',$ if there is an an active site in $q$ prior to applying $U_j',$ any element of $\mathcal{Q}_0$ contained in $q$ would have no active site with probability at most $\varepsilon_j.$ Thus, the probability having some $d\in \mathcal{Q}_0$ satisfying $d\subseteq q$ with no active sites is no larger than $e^{-\ell_0^d/6}+\varepsilon_j.$ Taking the union bound over the condition for all $q\in \mathcal{Q}_j$ satisfying $q\subseteq Q$, and removing the conditioning on $a\geq \vert Q_*\vert /8,$ we get
    \begin{align}
        \mathbf{P}\left[\exists D\in \mathcal{Q}_0, D\subseteq Q\colon \text{No active sites in }D\right] &\leq \varepsilon_j + e^{-\frac{\ell_j^d}{12}} + \left(\frac{\ell_{j+1}}{\ell_j}\right)^d\left[\varepsilon_j + e^{-\ell_0^d/6}\right]\nonumber\\
        &\leq \varepsilon_{j+1},
    \end{align} establishing the inductive step, and thus the claim.
\end{proof}
We have thus established that the hypercube of side-length $\ell_0$ containing the site $y$ we wish to spread the operator to contains at least one active site with probability $1-\varepsilon_k.$ To make this a probabilistic statement about an active site being on $y$ itself, we note that, by Lemma~\ref{lem:zz}, $W:=U_k\exp\left(-i\pi \frac{\ell_0^\alpha}{8}\sum_{q\in\mathcal{Q}_0}\ell_0^{-\alpha}H_\sigma^{(q)}\right)\mathcal{C}$ has probability $\frac{1-\varepsilon_k}{3}$ of propagating an active site to $y.$ Further, by the same logic as that which led to Eq.~(\ref{eq:Qover}), we have that $W$ results in at least $R^d/8$ active sites with a probability of at least $1-\varepsilon_k - e^{-R^d/12}.$

What remains to be shown is a choice of $\ell_j$s that is able to be implemented in time polylogarithmic in $R$ and such that $\varepsilon_k$ is small. We see that the time to implement $U_j$, denoted $T_j,$ assuming the Clifford layers are arbitrarily fast, satisfies the following recurrence: $T_{j+1} = 2T_j + t_j + \pi\ell_0^\alpha/8,$ with $T_0=0.$ We thus see that
\begin{align}
    T_j &= \sum_{l=0}^{j-1} 2^{j-1-l}t_l + (2^j-1)\frac{\pi\ell_0^\alpha}{8}\nonumber\\
    &\leq (2^j-1) \ell_0^\alpha\left(\frac{\pi}{8} + \sum_{l=0}^{j-1} \frac{\ell_{l+1}^\alpha}{2^l \ell_l^d}\right).
\end{align} If we choose $\ell_{l+1} \asymp \ell_l^{d/\alpha},$ we are able to get $T_j \lesssim 2^j\ell_0^\alpha.$ Thus, if $\ell_0$ is at most polylog in $R$, and $k$ is $\log\log R$, we would be able to get polylog run time.

We see that with this choice, $\log \ell_{j+1} \asymp \frac{d}{\alpha} \log \ell_j,$ and thus that $\log\ell_j \asymp \left(\frac{d}{\alpha}\right)^j\log \ell_0.$ Thus, we establish that \begin{equation}k\asymp\frac{\log\log R - \log \log \ell_0}{\log \frac{d}{\alpha}},\end{equation}
which, if $\ell_0$ is not too small, guarantees the appropriate $k$ will be $\log \log$ of $R.$ We next must guarantee that $\epsilon_k$ is sufficiently small. Note that $\varepsilon_k \leq 2^k R^d e^{-\ell_0^d/12}.$ Plugging in our expression for $k$, we get
\begin{align}
    \varepsilon_k \lesssim  R^d \log^{\frac{\log 2}{\log \frac{d}{\alpha}}}Re^{-\ell_0^d/12}.
\end{align} It thus suffices to take $\ell_0 \geq (13d)^{1/d}\log^{1/d}R$ to guarantee $\varepsilon_j = o(1).$ We thus see that $T_k\lesssim (\log R)^{\frac{\alpha}{d} + \frac{\log 2}{\log \frac{d}{\alpha}}}.$ Adding the final Clifford layer and applications of the normalized $H_\sigma^{(q)}$ for all $q\in\mathcal{Q}_0$ does not change that scaling.

For completeness, we note that some more work is needed to make the protocol work for every $R\in \mathbb{N},$ as not every $R$ can be written as as the product of $m_j$s which scale appropriately. However, we note that if $R$ is a power of $2,$ we can choose $\ell_0:=2^{\left\lceil\frac{\log_2(13d \log R)}{d}\right\rceil}$ and $m_{j+1}:=\min\left\{R/\ell_j, 2^{\left\lceil\left(\frac{d}{\alpha}-1\right)\log_2 \ell_j\right\rceil}\right\}$ to perform the protocol. For any $R$ not a power of $2,$ we can then embed two disjoint hypercubes $A$ and $B$ of sidelength $\mathfrak{L}$ into $Q_R,$ with $x\in A$ and $y\in B,$ where $\mathfrak{L}$ is a power of two with $\mathfrak{L}/R\in (1/4, 1/2).$ We can perform the protocol on $A,$ apply $R^{-\alpha} \sum_{(x,y)\in A\times B}\sigma_{xy} Z_xZ_y$ for time $1$ followed by another layer of random single-qubit Cliffords, and then perform the protocol on $B,$ and obtain a probability of at least $1/3-o(1)$ of successful operator spreading. To see this probability is the case, note that the protocol on $A$ gives at least $\mathfrak{L}^d/8$ active sites with probability $1-o(1),$ which the $ZZ$ pulse between $A$ and $B$ transmits to at least one active site in $B$ with a probability of at least $1-o(1)$ after the layer of random cliffords. The protocol on $B$ then produces an active site on $y$ with probability of at least $1/3-o(1),$ arriving at our lower bound.

\vspace{2mm}
\noindent
\textbf{Acknowledgments.}
This project used GPT-6 Astra for coming up with proof ideas and methods, and for checking and proofreading.
This work was supported in part by the DoE ASCR Quantum Testbed Pathfinder program (award No.~DE-SC0024220), NSF STAQ program, NQVL:QSTD:Design:FTL, NSF QLCI (award No.~OSI-2553574), AFOSR MURI, ONR MURI, and ARL (W911NF-24-2-0107). We also acknowledge support from the U.S.~Department of Energy, Office of Science, National Quantum Information Science Research Centers, Quantum Systems Accelerator (award No.~DE-SCL0000121) and from the U.S.~Department of Energy, Office of Science, Accelerated Research in Quantum Computing, Fundamental Algorithmic Research toward Quantum Utility (FAR-Qu). L.S. also acknowledges support through the NRC Research Associateship Program at the National Institute of Standards and Technology (NIST).

\bibliographystyle{amsalpha_edit}
\bibliography{lrb.bib}

\end{document}